\documentclass[12pt]{article}

\usepackage{setspace}

\usepackage{amsmath}
\usepackage{graphicx}
\usepackage{natbib}
\usepackage{url} % not crucial - just used below for the URL 
\usepackage{latexsym}
\usepackage{amsfonts}
\usepackage{amssymb}
\usepackage{psfrag}
\usepackage{graphicx}
\usepackage{calc, cancel}
\usepackage{graphics, graphicx, caption, subcaption}
\usepackage{latexsym, pgfplots, tikz, url,verbatim,multicol,enumerate,bm,listings,color}
\usepackage{multirow}
\usepackage[justification=centering]{caption}

\usepackage{wrapfig, sidecap,enumitem}
\usepackage{epsfig}
\usepackage{longtable}
\usepackage{hyperref}
\usepackage{graphicx}
\usepackage{natbib}
\usepackage{color}
\usepackage{amsgen,amsmath,amstext,amsbsy,amsopn,amssymb}
\usepackage{geometry}
\usepackage{subcaption}
\usepackage{caption}
\usepackage{wrapfig}
\usepackage{fancyhdr}
\usepackage{lastpage}
\usepackage{longtable}
\usepackage{enumitem}

\usepackage{amssymb}
\usepackage{xcolor}

\usepackage{amsmath}
\usepackage{graphics}%for scalebox
\usepackage{subcaption} %for subfigure

\DeclareMathOperator*{\argmax}{arg\,max}

\newcommand{\mSigma}{\boldsymbol{\Sigma}}
\newcommand{\vecmu}{\boldsymbol{\mu}}

\newcommand{\vectheta}{\boldsymbol{\theta}}

\newcommand{\veczero}{\mathbf{0}}
\newcommand{\vecone}{\mathbf{1}}
\newcommand{\ident}{\mathbf{I}}

\newcommand{\mata}{\mathbf{A}}

\newcommand{\mate}{\mathbf{E}}
\newcommand{\matd}{\mathbf{D}}
\newcommand{\matt}{\mathbf{T}}
\newcommand{\matr}{\mathbf{R}}
\newcommand{\mats}{\mathbf{S}}

\newcommand{\diag}{\,\mbox{diag}}
\newcommand{\tr}{\,\mbox{tr}}

\newcommand{\veca}{\mathbf{a}}
\newcommand{\vece}{\mathbf{e}}
\newcommand{\vecd}{\mathbf{d}}
\newcommand{\vecx}{\mathbf{x}}

\begin{document}
	\doublespacing

	\title{\bf Estimation of Gaussian Bi-Clusters with General Block-Diagonal Covariance Matrix and Applications}
	
	\author{Anastasiia Livochka \footnote{Department of Statistics and Actuarial Science, 200 University Ave W, Waterloo, ON, N2L 3G1, Canada} \and Ryan Browne \footnote{Department of Statistics and Actuarial Science, 200 University Ave W, Waterloo, ON, N2L 3G1, Canada. e: ryan.browne@uwaterloo.ca } \and Sanjeena Subedi\footnote{School of Mathematics and Statistics, 4302 Herzberg Laboratories, Carleton University, 1125 Colonel By Drive, Ottawa, ON, K1S 5B6, Canada e: sanjeena.dang@carleton.ca}}

	\maketitle

\begin{abstract}
Bi-clustering is a technique that allows for the simultaneous clustering of observations and features in a dataset. This technique is often used in bioinformatics, text mining, and time series analysis. An important advantage of biclustering algorithm is the ability to uncover multiple ``views'' (i.e., through rows and column groupings) in the data. Several Gaussian mixture model based biclustering approach currently exist in the literature. However, they impose severe restrictions on the structure of the covariance matrix. Here, we propose a Gaussian mixture model-based bi-clustering approach that provides a more flexible block-diagonal covariance structure. We show that the clustering accuracy of the proposed model is comparable to other known techniques but our approach provides a more flexible covariance structure and has substantially lower computational time. We demonstrate the application of the proposed model in bioinformatics and topic modelling.
\end{abstract}

\textbf{Keywords}: Bi-clustering, Gaussian mixture models

\section{Introduction}

Humans are particularly good at classifying objects into natural groups based on some notion of similarity and the goal they want to achieve. However, as the set of objects to organize and the number of features grows, we require faster and more memory-efficient tools to complete the task. Cluster analysis is an important area of unsupervised machine learning that deals with the problem of grouping objects so that objects in one cluster are more similar to each other than to objects in other clusters \citep{diday1976clustering}. The main applications of cluster analysis include data exploration and pattern recognition \citep{berkhin2006survey}. 

Cluster analysis has been widely studied for decades, and many clustering algorithms have been developed. Interestingly, the great variety of the available models can be attributed to the fact that precisely defining a cluster presents a problem \citep{milligan1987methodology}. Some algorithms (e.g. DBSCAN, HDBSCAN) leverage density assumptions by defining clusters as connected dense regions in the space \citep{khan2014dbscan, mcinnes2017hdbscan}. Other approaches view clusters as mixtures of certain distributions, with Gaussian mixture model (GMM) being the most popular one \citep{reynolds2009gaussian}. 

Bi-clustering is a technique that allows for the simultaneous clustering of observations and features in a dataset \citep{hartigan1972direct}. An important advantage of bi-clustering  is the ability to uncover multiple ``views" (i.e., through row and column groupings) in the data. This technique is often used in bioinformatics, text mining, and time series analysis \citep{madeira2004biclustering, madeira2008identification, busygin2008biclustering}. For example, \cite{wang2013} utilized biclustering algorithm on gene expression data and identified subgroups of breast cancer tumour with similar clinical characteristics. In topics modelling context, it's been extensively used to interpret the resulting clusters \citep{rugeles2017biclustering}.

There is no single definition of a cluster and as such there is no single definition of a bi-cluster. Therefore, a variety of bi-clustering models exist. One of the first applications of bi-clustering to text documents is based on the  spectral graph \citep{Dhillon2001} partitioning heuristic. Similarly, \cite{spectral_genes} shows that the checkerboard structure of the data matrix can be inferred from its eigenvectors. There are many greedy approaches proposed for the co-clustering of gene expressions \citep{padilha2017systematic,  cheng2000biclustering, bergmann2003iterative}. For instance, \cite{ben2002discovering} introduces an approach that finds the hidden order-preserving sub-matrices in the data matrix.

\begin{figure}[h!]
  \centering
{\includegraphics[width=1\linewidth]{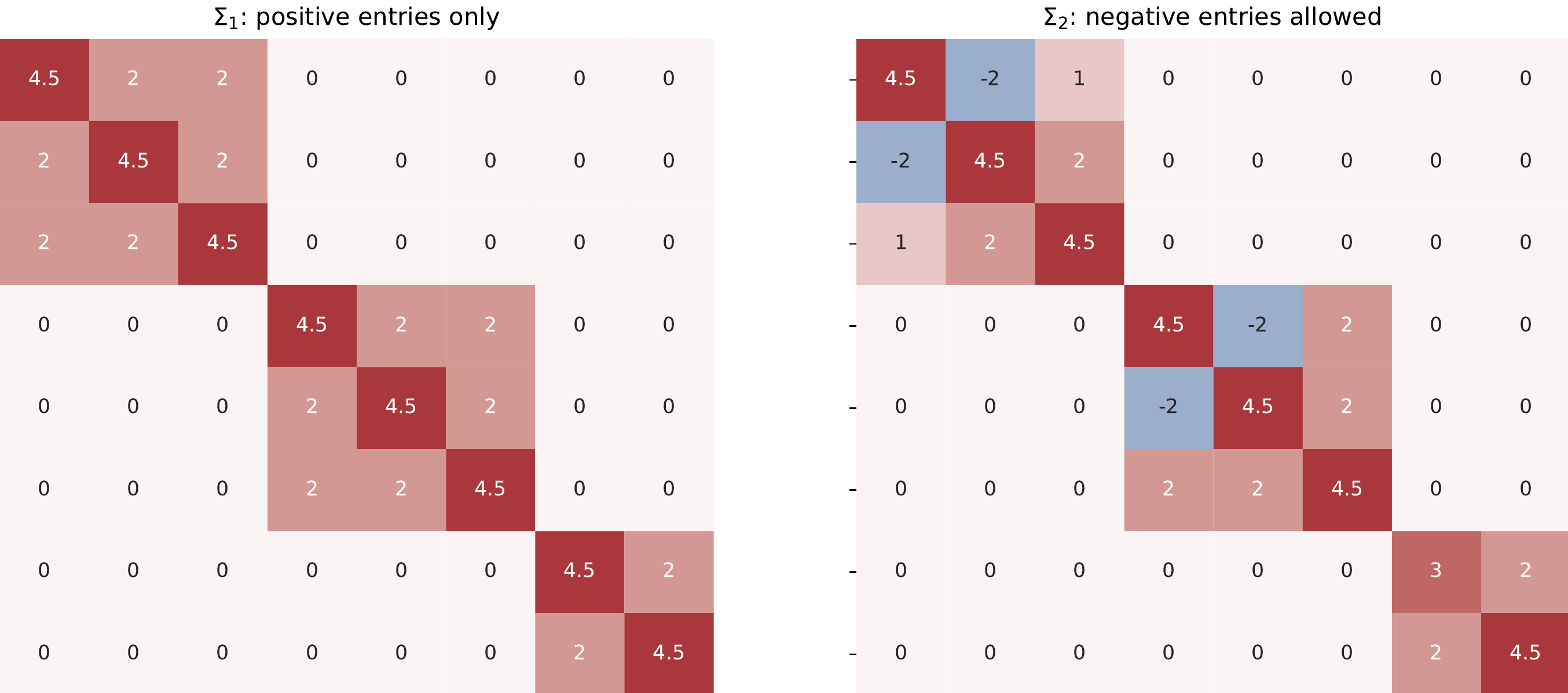}}\vspace{1mm}
  \caption{Illustration of the block diagonal matrices with off-diagonal elements limited to positive (\textit{left}) and general (\textit{right}) structure.}
  \label{fig:matrices_1}
\end{figure}

Alternate model-based biclustering approaches assume that the observations in the same row cluster were generated from the same distribution. e.g. $\mathcal{N}_p(\vecmu, \mSigma)$. In turn, column groups can be inferred from the structure of the covariance matrix. Suppose we have a $p$-dimensional mean vector $\vecmu$ such that
\[
\vecmu = \left[\vecmu_1, \vecmu_2, \vecmu_3 \right],
\]
where the dimesions of $\vecmu_i$ is $c_i$, $i=1,\ldots,3$ and $p=c_1+c_2+c_3$,
% such that
% \[
% \vecmu = \left[\overbrace{\mu_1,\ldots,\mu_{c_1}}^{\vecmu_1},\overbrace{\mu_{c_1+1},\ldots,\mu_{c_1+c_2}}^{\vecmu_2}, \overbrace{\mu_{c_1+c_2+1},\ldots,\mu_{c_1+c_2+c_3}}^{\vecmu_3} \right],
% \]
and covariance covariance matrix, $\mSigma$, can be written as:
\begin{equation}
\mSigma = 
\diag\left( \mSigma_1, \mSigma_2, \mSigma_3 \right) = 
\left[ \begin{array}{ccc}
 \mSigma_{1(c_1\times c_1)} & \veczero  & \veczero   \\
 \veczero &   \mSigma_{2(c_2\times c_2)}   & \veczero  \\
 \veczero &  \veczero &  \mSigma_{3(c_3\times c_3)}  
\end{array} \right].
% \qquad
% \vecmu = \left[ \vecmu_1, \vecmu_2, \vecmu_3 \right]
\end{equation}
Then  ($\vecmu_1$, $\mSigma_1), (\vecmu_2, \mSigma_2), (\vecmu_3, \mSigma_3$) define three distinct column groupings. Note that column permutations may be required to achieve a checkerboard-like structure as illustrated in \ref{fig:matrices_1}. This adds to the non-triviality of the estimation of block-diagonal covariance matrix.

Initial model-based bi-clustering approaches introduced additional latent variables to indicate column clusters \citep{blockcluster, gu2008bayesian}. \cite{Gallaugher2022} proposed a Gaussian model-based clustering for high-dimensional data. They partition columns of the dataset twice - once by means and another by variances. Others \citep{Martella2008BiclusteringOG,monica,Tu2022AFO} leveraged the bi-clustering framework rooted in the mixtures of factor analyzers \citep{McNicholas2008ParsimoniousGM} approach . The advantage of the factor analyzer structure is that it decreases the number of estimated parameters by assuming a latent low-dimensional representation of high-dimensional variables. However, this representation restricts the covariance matrix structure that can be recovered. For example, \cite{Martella2008BiclusteringOG} and \cite{monica} require the within block off-diagonal elements of the covariance matrix to $1$. \cite{Tu2022AFO} relaxed some constraints by characterizing covariance to allow the all of the off-diagonal entries of $i$-th covariance block to be equal to a positive constant. This is illustrated in the left panel of Figure \ref{fig:matrices_1}. In addition, the parameter estimation utilized in \cite{Martella2008BiclusteringOG, monica,Tu2022AFO} has limitations  due to the complexity of finding column groupings.

In this work, we aim to build on the model-based bi-clustering procedure by allowing the covariance matrix of the underlying mixture components to have a general block-diagonal structure as shown in the right panel of Figure \ref{fig:matrices_1}. We derive and compare three approaches to estimate the column groupings; greedy, convex relaxation and hierarchical. Empirically the hierarchical approach is shown to be the best approach in terms of accuracy and computational efficiency. We also compare the hierarchical approach to readily available covariance matrix estimators \citep{LEDOIT2004365, friedman2008sparse, Broto22}. The readily available approaches do not satisfy our desirable conditions for the covariance estimator, namely ($i$) enforcing the block-diagonal structure on the covariance matrix; and ($ii$) allowing for general covariance matrix structure inside the blocks. In the second half the manuscript, we introduce the model-based bi-clustering procedure with the general block-diagonal structure. We use simulated datasets to determine the efficacy of the model. We show utility of our approach in bioinformatics and topics modelling. First, we applied the proposed model and other existing bi-clustering models on two high-dimensional genomics classification problems: \texttt{Alon} \citep{alon} and \texttt{Golub} \citep{golub99} datasets. Then, we illustrate the proposed methodology on two datasets from topic modelling: hotel ratings \& reviews \citep{barkha_bansal_2018} and  Disneyland ratings \& reviews \citep{Chillar_2022}.

\section{Estimating a Block-Diagonal Covariance Matrix}
\label{approach}

This section discusses parameterization of a block-diagonal matrix and its estimation. We propose three methods for estimating the block-diagonal covariance matrix (as illustrated on the \ref{fig:matrices_1} - which is crucial for the GMM-based bi-clustering procedure given section \ref{chapter:model}. For simplicity, we derive an estimation for one component of GMM here. In Section \ref{chapter:model}, we extend the techniques for multi-component mixture model. The main aim of this work is to propose an approach such that: $(i)$ it imposes a block-diagonal structure onto the covariance matrix; $(ii)$ it does not restrict the covariance matrix structure inside the blocks. 

% As described in the literature review, many methods exist to estimate a general covariance matrix with a block-diagonal structure. However, there is no guarantee that real-world data will indeed be generated from a model with such covariance. Therefore, we need to impose this restriction artificially for bi-clustering purposes. 

\subsection{Imposing Block-Diagonal Structure Onto a Covariance Matrix}
Suppose we have a normally distributed data $\mathbf{X} \sim \mathcal{N}_p(\vecmu, \mSigma)$. One way to compute feature clusters, common for biclustering, is by estimating a block-diagonal covariance matrix that is close to the original $\mSigma$. In this  work, we propose to parameterize this block-diagonal covariance matrix as:
\begin{equation}     \label{eq:param}
     \matd_1^\intercal \mSigma \matd_1 + \dots + \matd_K^\intercal\mSigma \matd_K
\qquad  \mbox{where } \qquad
\matd_1 + \dots + \matd_K = \ident_p
\end{equation}
where, $K$ is the number of variable or column groups, $\matd_k = \diag(\vecd_k)$ and each element of $\vecd_k$ is zero or one.
In other words, the $\matd_k$ are binary diagonal matrices that sum up to the identity matrix. This allows for each $\matd_k$ to define a column or feature group $k$ using $\matd = \left[ \vecd_1, \ldots, \vecd_K \right]$. For example, to get the general covariance structure in Figure \ref{fig:matrices_1}, we set 
\begin{equation}   \label{example d}
\matd = 
\left[
\begin{array}{ccc}
  \vecd_1 & \vecd_2 & \vecd_3
\end{array}
\right]
=
\left[
\begin{array}{cccccccc}
 1 & 1 & 1 & 0 & 0 & 0 & 0 &  0  \\
 0 & 0 & 0 & 1 & 1 & 1 & 0 &  0   \\
 0 & 0 & 0 & 0 & 0 & 0 & 1 &  1   
\end{array}
\right]^\intercal .
\end{equation}
This proposed parametrization imposes a general block-diagonal structure on the covariance matrix. i.e.  we can obtain the  general covariance structure shown in the right panel of Figure \ref{fig:matrices_1}.

If we have a sample $\{\vecx_1, \vecx_2, \dots \vecx_N\}$ generated from a $p$-dimensional multivariate Gaussian with mean, $\vecmu$, and a postive definite  covariance matrix, $\mSigma$. The scaled or normalized log-likelihood is 
\begin{equation*}
    -\frac{1}{2} \log | \mSigma | - \frac{1}{2 N} \sum_{i=1}^N \tr\left[ \mSigma^{-1}( \vecx_i - \hat{\vecmu})  ( \vecx_i - \hat{\vecmu})^\intercal \right]. 
\end{equation*}
The maximum likelihood estimates (MLEs) of the parameters $\vecmu$ and $\mSigma$ are the sample mean and sample covariance:
\begin{equation*}
    \hat{\vecmu}  = \overline{\vecx} = \frac{\sum_{i=1}^N \vecx_i}{N},
    \qquad \mbox{and} \qquad 
   \mats = \frac{\sum_{i=1}^N (\vecx_i - \hat{\vecmu})(\vecx_i - \hat{\vecmu})^\intercal}{N}.
\end{equation*}
Furthermore, we denote sample correlation matrix by $\mats^{\star}$. 

Since, the MLE for unconstrained covariance matrix is  $\mats$, the log-likelihood function for the block diagonal covariance matrix as a function of the feature groups, $\matd = [ \vecd_1, \dots \vecd_K]$, is
\begin{equation} \label{obj fn}
\mathcal{L}(\vecd_1, \ldots, \vecd_K )  
 =  -\frac{1}{2} \log \left| \sum_{k=1}^K \matd_k \mats \matd_k \right| - \frac{1}{2} \tr\left[  \left( \sum_{k=1}^K \matd_k \mats \matd_k \right)^{-1} \mats \right] 
\end{equation}
where $\matd_k = \diag(\vecd_k)$.  In the next three sections, we discuss three approaches to estimate $\matd$: greedy, convex relaxation (numerical) and hierarchical. 

\subsection{Greedy Approach}
Motivated by the approach in \cite{Tu2022AFO}, we decided to adopt a greedy algorithm for estimating $\matd = [ \vecd_1, \dots \vecd_K]$ in (\ref{obj fn}).  To initialize we follow \cite{Tu2022AFO} and set $\matd$ to the first $K$ principal components of the correlation matrix $\mats^\star$. Then, in the $j^{th}$ row of $\matd$, we find the largest element, e.g. the $k^{th}$  element, set that element to one and then the remaining are set zero. This can be interpreted as assigning the $j$-th row to the $k^{th}$-the column group. 

% Then for each row of $\matd$, we find the largest element, set that element to one and then the remaining are set zero.  This can be interpreted as assigning the features or variables to the  column group.  \textcolor{blue} {which is better?  }
%Then for the  $j^{th}$ column we compute $p = argmax(P_j)$, where $P_j$ denotes $j$-th column of $P$. 

After initialization, we iteratively reassign column clusters similar to the initialization but instead we use the largest log-likelihood value (\ref{obj fn}). Starting with variable $j$, we evaluate  the equation (\ref{obj fn}) by considering this feature or variable in each of the $K$ feature group and assign the variable to the $k^{th}$ column group that maximizes the log-likelihood. Alternatively, this is equivalent to considering each of the unit vectors,  of length $K$ in  the row $j^{th}$ of $\matd$. Mathematically, for the $j^{th}$ row  of $\matd$, denoted by $\matd_{j}$, we have
\begin{equation*}
    \matd_{j} = \argmax_{ \veca \in \{ \vece_1, \ldots,  \vece_K \} } 
    \mathcal{L}(\vecd_1, \ldots, \vecd_{j-1}, \veca, \vecd_{j+1} \ldots, \vecd_K )
\end{equation*}
for $j = 1,\ldots, p$ and $\vece_k$ is a unit vector of length $K$ which has a unit in the $k^{th}$ position and zero elsewhere. 
%Starting with the first column $j = 1$, we choose: 
%\begin{equation*}
%    p = argmax_{1 \leq k \leq K} \mathcal{L}(\hat{\mTheta}_k), \text{ with } \hat{\mTheta}_k = \sum_{i=1}^K \matd_i\hat{\mSigma} \matd_i, \text{  } \matd_{i = k}[j, j] = 1 \land \matd_{i \neq k}[j, j] = 0
%\end{equation*}
%The cluster assignment for the $j$-th row is updated by setting $\matd_p[j, j] = 1$ and $D_{k \neq p}[j, j] = 0$. 
This means each updated row is conditional on the rest of them.
%Conditioning on the first updated row, we repeat the procedure for the rest of them. 

\subsection{Convex Relaxation Approach}

We aim to find $\matd = \left[\vecd_1, \vecd_2, \dots, \vecd_K \right]$ that maximizes log-likelihood in (\ref{obj fn}). Keeping in mind that $\vecd_k$ should be binary by definition of  (\ref{eq:param}), this is  a non-linear integer problem, which is known for its complexity \citep{Hemmecke_2009}. To simplify, we relaxed the $\vecd$ binary constraints and use quadratic programming to solve the following:
\begin{equation}
    \max \mathcal{L} \quad \text{st.} \quad \forall k: \ \vecd_k > \veczero \quad \text{and} \quad \sum_{k=1}^K \vecd_k = \vecone 
\end{equation} 
where $\vecd_k > \veczero$ means that each element of $\vecd_k$ should be greater than zero. This relaxes the condition that the $\matd_k$ are binary diagonal matrices by allow each element of $\matd_k$ to be positive. A possible  solution is that $\forall k: \ \vecd_k = 1/K \times \vecone$  as it results in $\mSigma = \mats$. To avoid the trivial solution, we added a penalty on the logarithm of the determinant of $\matd^\intercal \matd$. Then in addition we will use a penalty to push $\vecd_k$ towards binary values. In particular the penalty is $\sum_{k = 1}^K \vecd_k^\intercal\vecd_k - p$. 
Finally, instead of applying a block diagonal structure to the covariance, we apply it to the inverse of the covariance matrix also known as the precision matrix. Note that the two parameterizations are equivalent because the inverse of a block diagonal matrix is block diagonal matrix. However, using the precision matrix will simply the derivatives.

The parameterized block diagonal precision matrix is
\begin{equation}  \label{theta-conv}
   \matd_1 \mSigma^{-1} \matd_1 + \dots + \matd_K \mSigma^{-1} \matd_K.
\end{equation}
We want to maximize the log-likelihood function with respect to the block-diagonal structure of the covariance matrix:
The MLE for $\mSigma^{-1}$ is $\mats^{-1}$ so putting (\ref{obj fn}) in terms of the precision matrix we have 
\begin{equation}  \label{ll-star-1}
\mathcal{L}^{\star} (\vecd_1, \ldots, \vecd_K )  
  \frac{1}{2} \log \left| \sum_{k=1}^K \matd_k \mats^{-1} \matd_k \right| - \frac{1}{2} \tr\left[  \left( \sum_{k=1}^K \matd_k \mats^{-1} \matd_k \right) \mats \right] 
\end{equation}
We utilize the properties of the trace and element-wise product operators to obtain
\begin{equation}   \label{trace-prop}
    \tr(\matd_k \mats^{-1} \matd_k \mats) = \vecd_k^{T} \left(\mats^{-1} \odot\mats \right) \vecd_k
\end{equation}
where $\odot$ is element-wise or Hadamard product. 

Combing the penalties and the log-likelihood function we have
\begin{equation} 
\mathcal{F}^{\star} (\vecd_1, \ldots, \vecd_K )  
= \mathcal{L}^{\star} (\vecd_1, \ldots, \vecd_K )   + \gamma \left(\sum_{k = 1}^K \vecd_k^\intercal\vecd_k - m \right) - \lambda \log \left| \matd^\intercal \matd \right|
    \label{final-Q}
\end{equation}
for $\gamma, \lambda \in  \mathbb{R}^{+}$. This optimization problem can be solved numerically utilizing gradient descent. Gradient descent is an iterative, first-order optimization method that finds local minima or maxima of the differentiable function. We propose to estimate the columns of $\matd$ iteratively conditioning on the values of all other column vectors.

To compute the partial derivatives of (\ref{final-Q}) with respect to $\vecd_k$, we note that $\matd_k = \diag(\vecd_k)$ and $\partial \diag(\vecd_k) = \diag(\partial \vecd_k)$. The partial derivative is
% \begin{equation}
%    \frac{\partial }{\partial \vecd_k}  \frac{1}{2} \log \left|\sum_{k=1}^K \matd_k \mats^{-1} \matd_k \right| = \left[ \matt \odot \mats^{-1} \right] \vecd_k
%    \label{partial-log}
%\end{equation}
%We utilize the term rearrangement suggested in equation (\ref{trace-prop}):
% \begin{equation}
%   \frac{\partial}{\partial \vecd_k} \left[ - \frac{1}{2 }\tr\left(\sum_{k=1}^K \matd_k \mats^{-1} \matd_k  \mats \right) \right] 
%   = - \left(\matr^{-1} \odot \matr \right) \vecd_k
%    \label{partial-main}
%\end{equation}
% \begin{equation}
%    \frac{\partial }{\partial \vecd_k}  \mathcal{L}^{\star} 
%    = \left[ \matt \odot \mats^{-1} \right] \vecd_k - \left(\matr^{-1} \odot \matr \right) \vecd_k
%    \label{partial-log}
%\end{equation}
%Where $\matt = (\sum_{k=1}^K \matd_k \mats^{-1} \matd_k)^{-1}$. 
 \begin{equation}
    \frac{\partial   \mathcal{F}^{\star}  }{\partial \vecd_k} 
    = \left[ \matt \odot \mats^{-1} \right] \vecd_k - \left(\matr^{-1} \odot \matr \right) \vecd_k
   + 2 \gamma \vecd_k  -2\lambda \left[ \matd(\matd^\intercal \matd)^{-1} \right]_k
    \label{partial-log}
\end{equation}
where $\matt = (\sum_{k=1}^K \matd_k \mats^{-1} \matd_k)^{-1}$ and $\left[\mata \right]_k$ denotes the $k$-the column of the matrix $\mata$.
% \begin{equation}
%    \frac{\partial}{\partial \vecd_k} (\gamma(\sum_{k = 1}^K \vecd_k^\intercal\vecd_k - m)) = 2 \gamma \vecd_k 
%    \label{partial-norm}
%\end{equation}
% \begin{equation}
%    \frac{\partial}{\partial \vecd_k} (- \lambda \log |\matd^\intercal \matd|) = -2\lambda(\matd(\matd^\intercal \matd)^{-1})_k
%    \label{partial-log-det}
%\end{equation}
%where $A_k$ denotes the $k$-th column of $A$. Combining all of the above derivations:
%\begin{equation}
%    \frac{\partial \mathcal{Q}}{\partial \vecd_k} = (\matt \odot \matr^{-1}) \vecd_k - (\matr^{-1} \odot \matr) \vecd_k + 2 \gamma \vecd_k  - 2\lambda(\matd(\matd^\intercal \matd)^{-1})_k
%\end{equation}
To maximize the $\mathcal{F}^{\star}$, we take a step in the direction of a positive gradient. Let us denote $\omega \in \mathbb{R}$ to be a learning rate, then the proposed update for $\vecd_k$ is defined as:
\begin{equation}
    \vecd_k^{\ (t + 1)} = \vecd_k^{\ (t)} + \omega \frac{\partial   \mathcal{F}^{\star}  }{\partial \vecd_k}. 
    \end{equation}
Through some experimentation, setting $\omega=0.1$ worked well in all the simulations.

\subsection{Hierarchical Clustering Approach}
\label{section:hierarchical}

Hierarchical clustering refers to a widely utilized family of clustering algorithms \citep{murtagh2014ward, beeferman2000agglomerative, erica2018agglomerative}, that are based on an iterative procedure of either merging or splitting nested clusters. Merging or splitting is also known as bottom-up and top-down approaches correspondingly. Many readily available implementations and generalizations of the algorithm ensure robustness on various input data configurations \citep{balcan2014robust}. 

Here, we leverage the bottom-up hierarchical approach known as agglomerative clustering to find the column clusters. However, instead of clustering the transpose of the dataset, we propose to use the absolute values of correlations as a feature representations for each variable. The absolute value will remove the effect of positive and negative correlations. Then we apply agglomerative clustering to the feature representations.
%$(i)$ similarity/dissimilarity metric between variables or $(ii)$ feature representations to cluster. 
To obtain the variables features, if $\mats^{\star}$  is theestimate of the correlation matrix based on dataset and $\mats$ is the MLE of the covariance matrix, we calculate
%\begin{equation*}
%    \mats^{\star} = \diag( \mats )^{-\frac{1}{2}} \ \mats \ \diag(\mats)^{-\frac{1}{2}}
%\end{equation*}
\begin{equation}
    \matr^\star = \left| \mats^\star  \right|
    =\left| \diag( \mats )^{-\frac{1}{2}} \ \mats \ \diag(\mats)^{-\frac{1}{2}}  \right|
    \label{R-hierarchical}
\end{equation}
where  $\left| \mata \right|$ applies the absolute value element-wise to the matrix $\mata$ and $\diag(\mata)$ creates a diagonal matrix using the diagonal elements of the matrix $\mata$. Treating $\matr^\star$ as feature vectors, we then use the $l_2$-norm to construct a distance matrix. This distance results in variables with similar correlation vectors to be close to each other.

The agglomerative clustering procedure starts by assigning every variable into its own group. On every iteration, it merges two of the most similar groups, as measured by a selected similarity/dissimilarity metric, which is known as linkage criteria. The linkage criterion is a metric that defines the distance between two clusters. There are three most widely used linkage criteria \citep{8862232}: single linkage, complete linkage and average linkage. 
single, complete and average linkage defines distance between two clusters as the minimum, maximum or average distance among distances between the two clusters, respectively. 
%Complete linkage defines distance between two clusters as the maximum distance among distances between the two clusters. The average linkage defines distance between two clusters as the average of the distance among distances between the two clusters. Single and complete linkage 
We propose utilizing average linkage criteria as this method results in clusters with the highest cohesion as discussed in \cite{Sokal1958ASM}. Once the $K$ variable groups have been obtained, we construct the corresponding block diagonal covariance matrix. 

%======================================================================
\subsection{Comparison \& Experiments}
%======================================================================
In this subsection, we aim to conduct experiments with simulated data to establish which of the three proposed approaches for estimating the block-diagonal matrix is the most accurate and efficient. Section \ref{Section:approaches} compares the execution time and accuracy of the three approaches while varying the number of observations in the dataset. We find that the hieracherical approach is the the most accurate and efficient one.

   The we compare the  hierarchical approach to other known covariance estimators. In the \ref{section:cov_comp_lit} we experiment with recovering two different structures of the covariance matrix with our vs state-of-the-art approaches, including the shrinkage estimator \citep{LEDOIT2004365}, Graphical Lasso \citep{friedman2008sparse}, factor-analyzer UCUU \citep{Tu2022AFO} and MLE. We find that methods that artificially enforce block-diagonal covariance structure are more accurate and the proposed method is indeed more robust than the UCUU model \citep{Tu2022AFO} for matrices with negative covariances.

Finally, we investigate the limitations of the hierarchical estimator by increasing the complexity the block-diagonal matrix $\mSigma$ when the number of columns group is known and unknown. We examine using the maximum silhouette score \citep{ROUSSEEUW198753} to determine the number of groups. We discovered that the usage of the hierarchical algorithm is limited when estimating high number of small blocks.

\subsubsection{Accuracy and Efficiency of the three Proposed Estimation Approaches}
\label{Section:approaches}
In this subsection, we investigate the accuracy and efficiency of the three proposed approaches to the problem of recovering column clusters $\matd_k$. The experiment setup is similar to other proposed in  \cite{Tu2022AFO}. In this subsection, we generate $N$ observations from $p=8$ dimensional Guassian distribution with $\vecmu = (0, 1, 2, 3, 4, 5, 6, 7)^\intercal$ and two different covariance matrices. Figure \ref{fig:matrices_1} displays two options for the covariance matrix that are considered in the experiments: $\mSigma_A$ with only non-negative off-diagonal entries; $\mSigma_B$ with negative off-diagonal entries. $\mSigma_A$ and $\mSigma_B$ are both block diagonal matrices displayed in Figure \ref{fig:matrices_1}. For case A, we have $\mSigma_A= \diag( \mSigma_{A1} ,  \mSigma_{A2} ,  \mSigma_{A3} )$  where 
\begin{equation*} 
 \mSigma_{A1} 
 = \left[ \begin{array}{ccc}
4.5 & 2  & 2   \\
2 &  4.5  & 2  \\
 2 &  2 &  4.5
\end{array} \right],
\quad
 \mSigma_{A2} 
 = \left[ \begin{array}{ccc}
4.5 & 2  & 2   \\
2 &  4.5  & 2  \\
 2 &  2 &  4.5
\end{array} \right],
\quad
 \mSigma_{A3} 
 = \left[ \begin{array}{cc}
4.5 & 2    \\
2 &  4.5    \\
\end{array} \right],
\end{equation*}
and for case B, we have $\mSigma_B= \diag( \mSigma_{B1},  \mSigma_{B2},  \mSigma_{B3} )$ where
\begin{equation*} 
\mSigma_{B1} 
 = \left[ \begin{array}{ccc}
4.5 & -2  & 1   \\
-2 &  4.5  & 2  \\
 1 &  2 &  4.5
\end{array} \right],
\quad
 \mSigma_{B2} 
 = \left[ \begin{array}{ccc}
4.5 & -2  & 2   \\
-2 &  4.5  & 2  \\
 2 &  2 &  4.5
\end{array} \right],
\quad
 \mSigma_{B3} 
 = \left[ \begin{array}{cc}
3 & 2    \\
2 &  4.5    \\
\end{array} \right].
\end{equation*}

The choice of covariance matrices is motivated by restrictions for off-diagonal elements to be strictly non-negative utilized commonly in the literature  \citep{monica, Tu2022AFO, Martella2008BiclusteringOG}. To goal here is to recover a covariance matrix of the general structure, including the possibility of negative off-diagonal elements. Note that both matrices in Figure \ref{fig:matrices_1} have a natural block-diagonal structure, which provides us with ground truth for column cluster indicator matrices $\matd_k$ as  shown previously in  (\ref{example d}). %For example, for both $\mSigma_1$ and $\mSigma_2$ we have $\matd_1 = \diag\left( 1, 1, 1, 0,  0, 0, 0, 0\right)$. 

\begin{figure}[t!]
\centering
\includegraphics[width=0.48\linewidth]{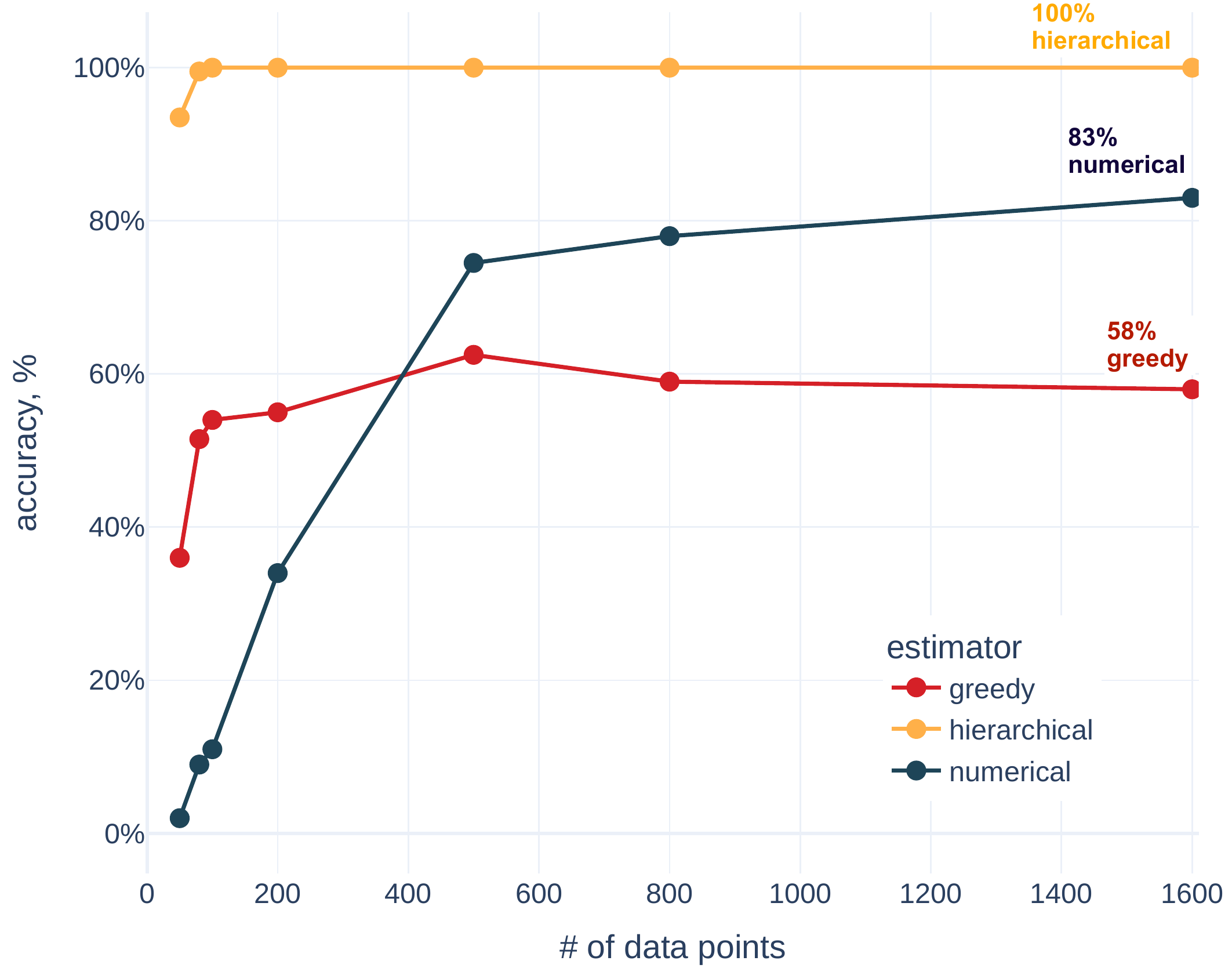} 
\includegraphics[width=0.48\linewidth]{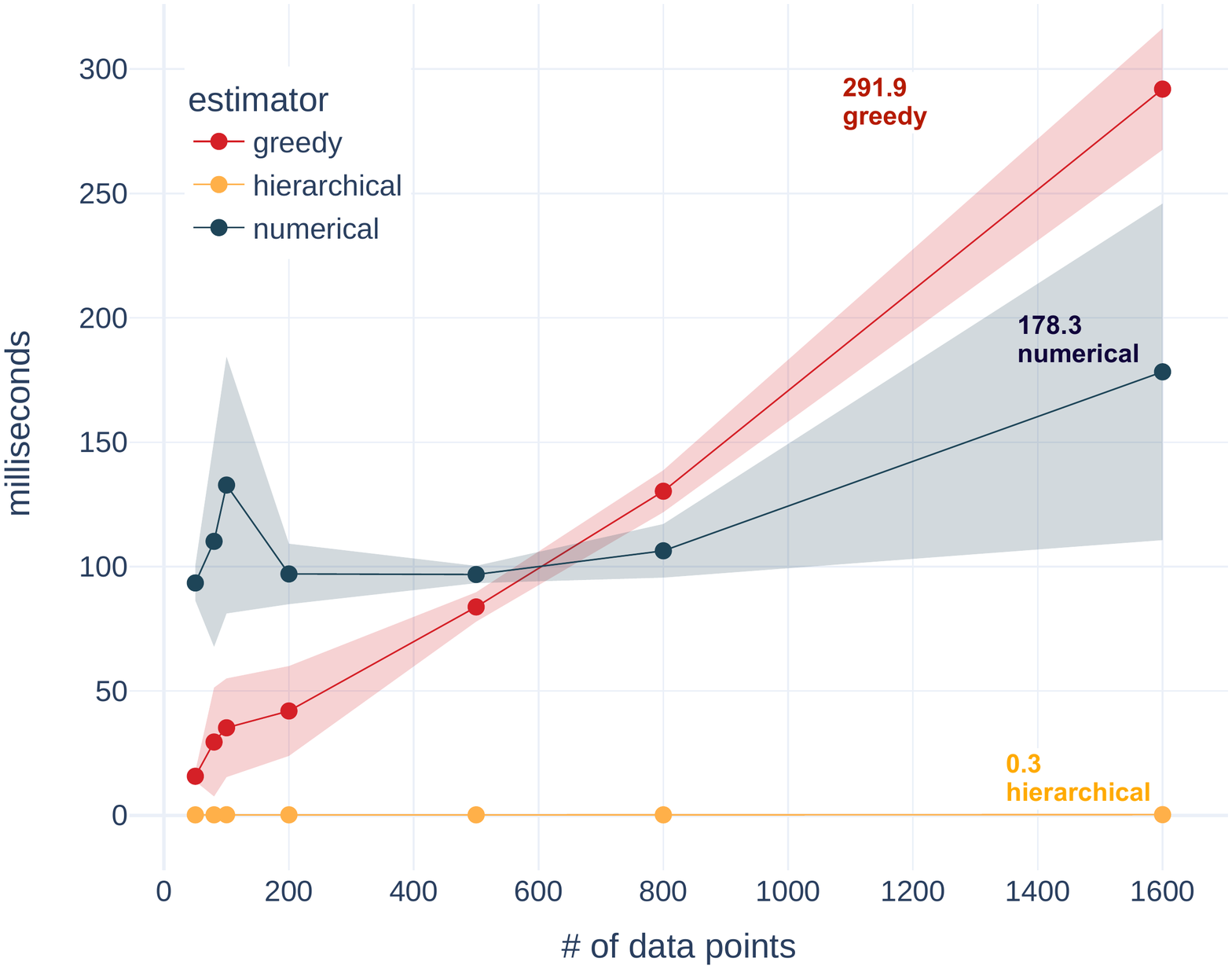}
% \captionsetup{justification=centering,margin=2cm}
\caption{ Comparison of accuracy \& execution time over $200$ runs for estimating $D_k$ for block-diagonal $8\times 8$ covariance matrix with $3$ blocks defined by $\mSigma_A$ which has only positive entries. ({\em Left}) The accuracy of the proposed approaches is given 
a number of the data points in the generated dataset $X$.
({\em Right}) Execution time of the proposed approaches.
}
\label{fig:exp_pos}
\end{figure}

Figure \ref{fig:exp_pos} shows results with data generated from $\mathcal{N}(\vecmu, \mSigma_A)$. We vary the number of data points, $N=50,80,100,200,500,800,1600$, to understand the impact on both accuracy and execution time of the proposed methods. For every size $N$, we run the experiment $200$ times and set number of column clusters to three. The accuracy is calculated as the total number of times when an algorithm returned correct column groups $\matd$ divided by the total number of attempts. \ref{fig:exp_pos} (\textit{right}) demonstrates the average execution time in milliseconds over $200$ runs together with it's standard deviation. 

From Figure \ref{fig:exp_pos}, it is clear that the hierarchical estimator for the column groups $\matd_k$ is both accurate and efficient. It gets to $100$\% starting from $100$ data points in the training set and demonstrates solid accuracy $> 90\%$ even with $N=50$. The numerical estimator benefits from the increasing dataset size. However, the time it needs to converge grows as well. In this experiment, the greedy estimator proved to be the least efficient and the least accurate for larger sample sizes.

\begin{figure}[h!]
\centering
\includegraphics[width=0.48\linewidth]{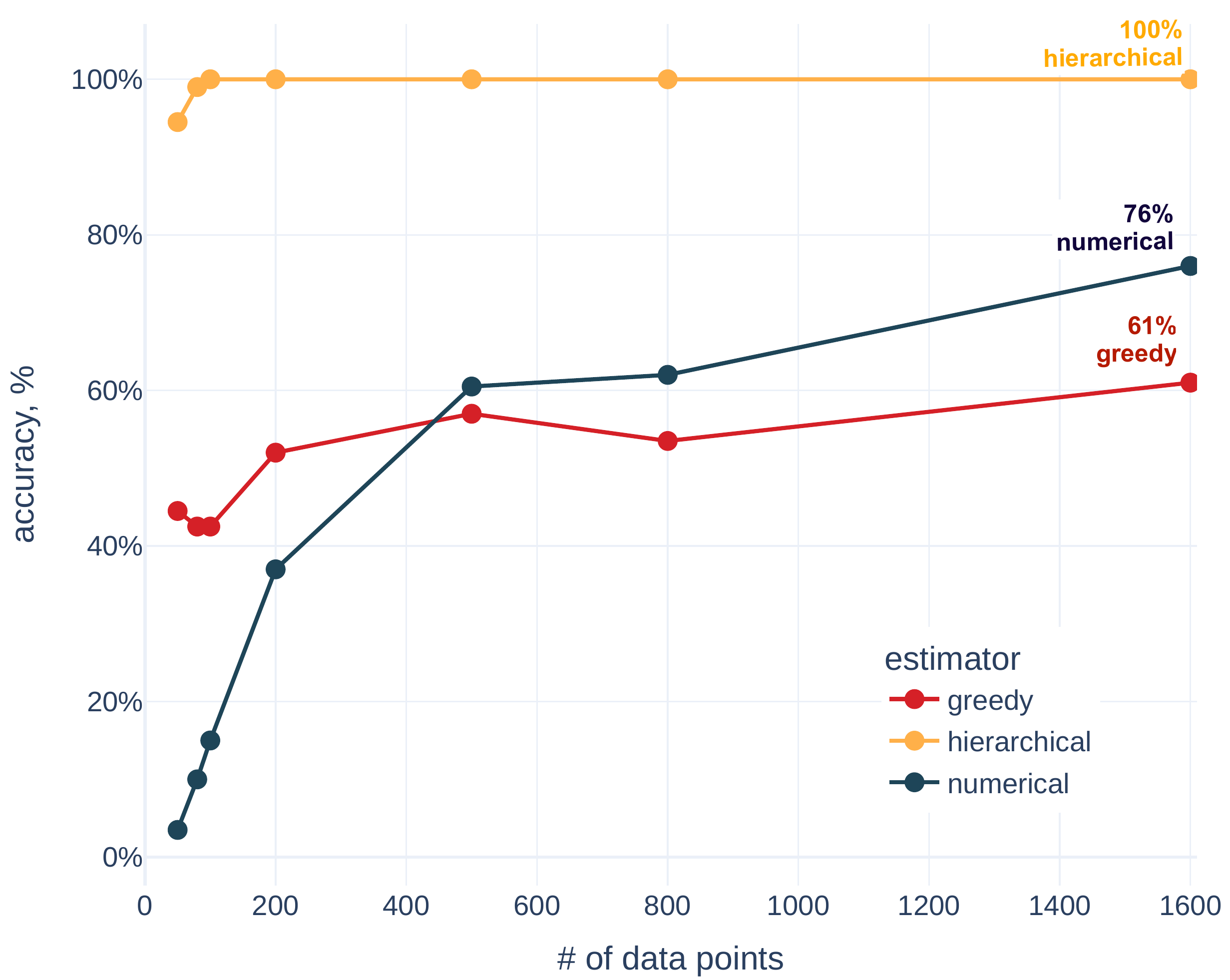} 
\includegraphics[width=0.465\linewidth]{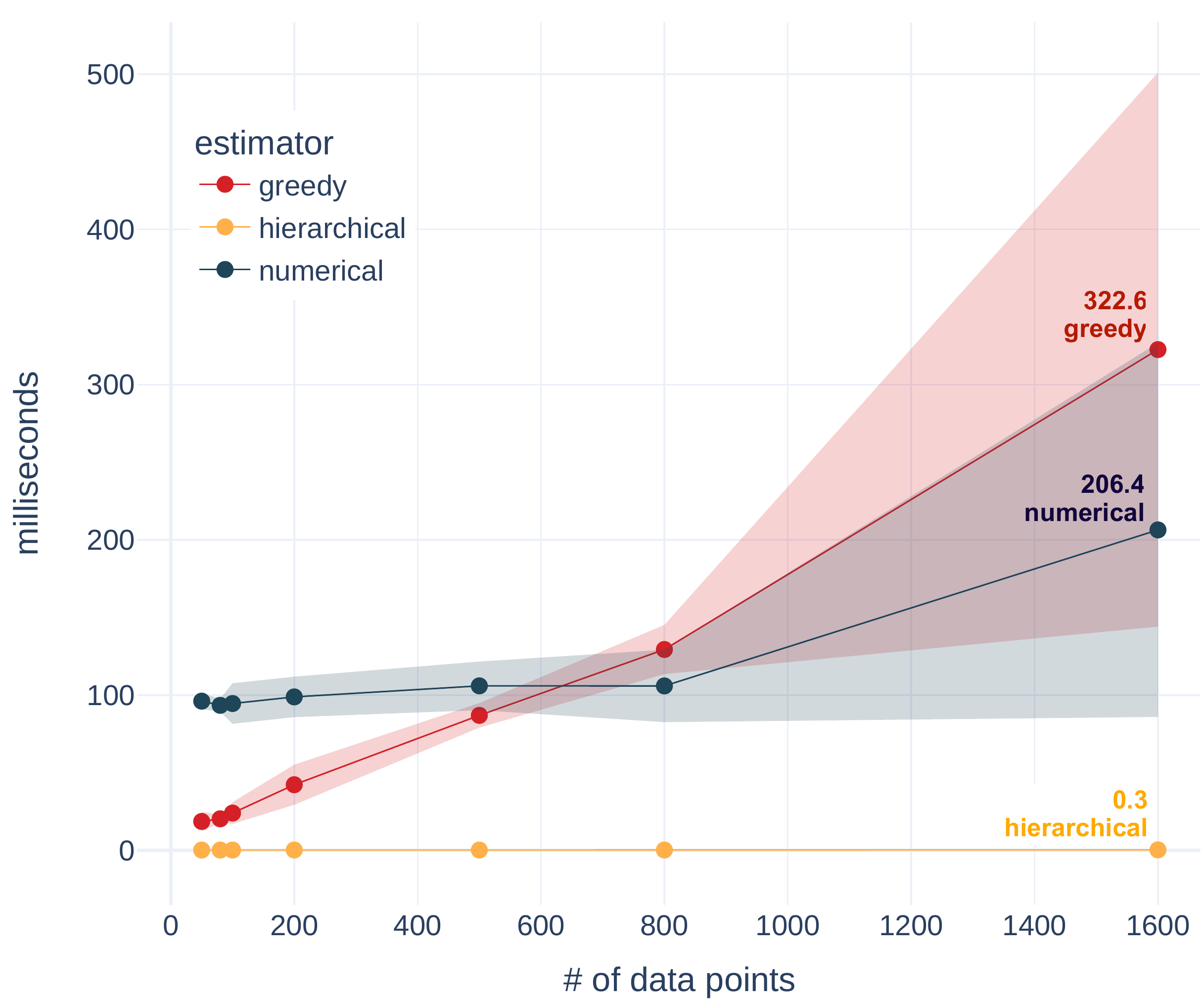}
% \captionsetup{justification=centering,margin=2cm}
\caption{ Comparison of accuracy \& execution time for estimating $D_k$ for block-diagonal $8\times 8$ covariance matrix with $3$ blocks defined by $\mSigma_B$ which has some negative entries. ({\em Left}) The accuracy of the proposed approaches given 
a number of the data points in the generated dataset $X$.
({\em Right}) Execution time of the proposed approaches.
}
\label{fig:exp_neg}
\end{figure}
Figure \ref{fig:exp_neg} demonstrates the results of the analogous experiment with data generated from $\mathcal{N}(\vecmu, \mSigma_B)$.  Interestingly, the execution time of both numerical and greedy estimators increases. While the numerical estimator demonstrates degraded accuracy, the greedy one improves in the presence of negative covariances or a greater variety of values within a covariance block. At the same time, the hierarchical approach leads in terms of both efficiency and accuracy, and it can correctly cluster columns even for the matrix with negative covariances. Therefore, for the following experiments, we proceed with the hierarchical approach.
 
 \subsubsection{Recovering Block-Diagonal Covariance Matrix \& Comparison with the Literature}
\label{section:cov_comp_lit}

%In the previous experiment, we focused on the detection accuracy of the feature or variables groups within the block-diagonal covariance matrix $\mSigma$. 

In this section, we evaluate the accuracy of estimation of the block-diagonal covariance matrix $\mSigma$ using the mean absolute percentage error (MAPE). For estimating $\mSigma$ with an estimate $\hat{\mSigma}$, we compute MAPE \% measure as following:
\begin{equation}
     100 \times \frac{\sum_{i, j}|\sigma_{ij} - \hat{\sigma}_{ij}|}{\sum_{i, j}\sigma_{ij}}.
    \label{mape}
\end{equation}
We compare the hierarchical approach to the MLE and three other methods utilized in the literature:  shrinkage estimator \citep{LEDOIT2004365}, Graphical Lasso \citep{friedman2008sparse} and UCUU model \citep{Tu2022AFO}.

\begin{figure}[h!]
  \centering
  \includegraphics[width=1\linewidth]{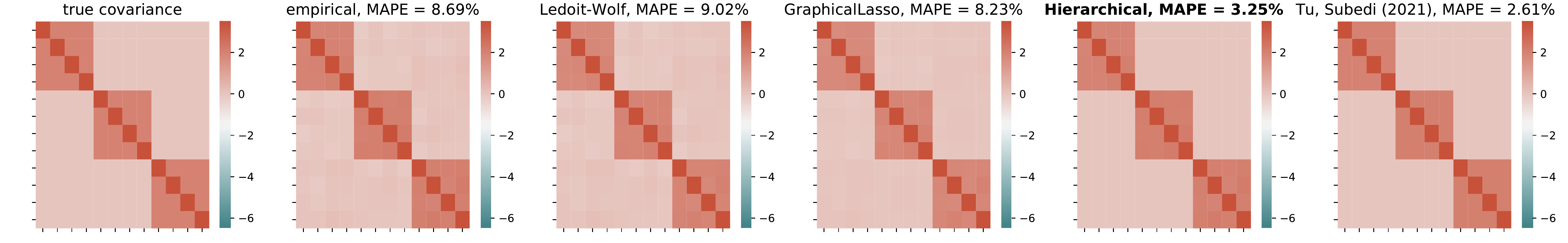} 
\includegraphics[width=1\linewidth]{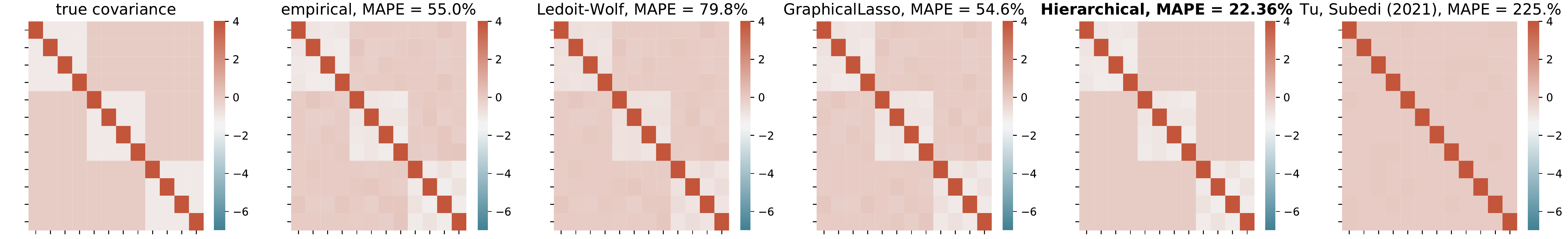}
  \caption{Average estimations over ten runs of recovering block-diagonal covariance matrix with only positive (\textit{top}) / negative (\textit{bottom}) off-diagonal entries with five methods including MLE (empirical), Ledoit-Wolf, Graphical Lasso, Hierarchical (proposed) \& UCUU model by Tu and Subedi (2021). For standard deviations, see \ref{appendix-a}.}
  \label{fig:comparison}
\end{figure}

The first column of Figure \ref{fig:comparison} shows two block-diagonal covariance matrices used to generate data-sets. The top row has a block-diagonal covariance matrix with $4.5$ on the diagonal and $2$ on the off-diagonals within a block. The bottom row has a block-diagonal covariance matrix with $4.5$ on the diagonal and $-1$ on the off-diagonals within a block. We use $N=300$, $p=12$ and $\vecmu=(1, 2, \ldots, 12)$. Every estimation of the covariance matrix is replicated ten times with the average $\hat{\Sigma}$ for each estimator depicted in Figure \ref{fig:comparison}.  For the hierarchical approach and UCUU model \citep{Tu2022AFO} we set the number of column groups to three. 

Our analysis show that artificially enforcing block-diagonal covariance structure is justified for the problems when such covariance structure is expected. Note that hierarchical approach and UCUU model are at least two times more accurate (in terms of MAPE) than others when the off-diagonal elements in the blocks are positive. On the other hand, the proposed hierarchical method is robust in the presence negative off-diagonal covariances, while the performance of the UCUU model by \cite{Tu2022AFO} suffers dramatically as it is not designed for such case.

Note that, MAPE is only helpful for relative algorithm accuracy comparison but not for absolute estimation quality assessment. It allows the comparisons of the form ``method X is 2 times more accurate than Y". However, we do not deduce that all methods are highly inaccurate for the covariance matrix with negative off-diagonal entries because the denominator of equation \ref{mape} is small. Therefore even small errors will result in large percentages. 

\subsection{Impact of the Covariance Block Structure on the Performance of Hierarchical Estimator}
\label{section:cov_structure}
%The previous sections concluded that the hierarchical estimator of column groups $\matd$ is the most accurate and efficient among the three proposed ones. It quickly reached close to $100$\% accuracy, even with small sample sizes (see \ref{fig:exp_pos}). 

In this section, we investigate the limitations of the hierarchical estimator by increasing the complexity the block-diagonal matrix $\mSigma$  when the number of columns group is known and unknown.  We measure the accuracy of the estimated column groups compared the generated column groups. The complexity of the variable groups $\matd$ is influenced by the dimensionality of number of features variables (i.e. columns) in the dataset, $p$; and the number of blocks or column groups, $K$. We fix the number of rows in the dataset to $N = 50$ and vary $p$ and $K$ to understand its impact on the estimator's accuracy. This choice is motivated by common real-life scenarios (e.g. genomics, natural language processing etc.) when the number of observations is often limited compared to the number of potential features. 

%\begin{table}[h!]
% \small
%    \centering
%    \begin{tabular}{c|c c c} 
%        level   
%            & low$=0$
%            & medium$=1$
%            & high$=2$\\ \hline
% \# variables   -    $p$ 
%        & $24$ & $96$  & $384$  \\
% \# features    -    K
%        & $3 \times 2^{\text{variable level}} $ & $4 \times 2^{\text{variable level}}$ & $8 \times 2^{\text{variable level}} $ \\
%\# observations  -      N
%        & $50$ & $50$ & $50$ \\
%    \end{tabular}
%\caption{Summary of different combinations of the number of variables, $p$, and the number of variable groups, $K$, in $\mSigma$. Note We set $K$ dynamically to $p$ (e.g. low number of blocks is $3$ for the data with $24$ variables and $12$ for the high-dimensional data with $384$ columns)}
%\label{table:exp2-setup}
%\end{table}

We evaluate the accuracy of the hierarchical estimator with three different levels for the number of variables, $p$, and the number of blocks or variable groups, $K$. They will be referenced as low, medium and high and summarized in  Table \ref{table:exp2-setup}. The number of column groups at every level is proportional to number of variables. For example, for the low-low setting $(p,K) = (24,3)$ while the medium-low setting has $(p,K) = (96,4)$. For simplicity, the number of variables within each block is set to $p/K$.

\begin{table}[h!]
\footnotesize
    \centering
    \begin{tabular}{c| l l l}  
$(p,K)$  &    \multicolumn{3}{c}{\# of variables - $p$}  \\    
 \# of column groups - $K$    &  low  & medium & high  \\ \hline
low     	& $(24,3)$		& $(96,4)$ 	& $(384, 8)$ \\
medium	& $(24,6)$		& $(96,8)$		& $(384,16)$ \\
high		& $(24,12)$	& $(96,16)$	& $(384,32)$ \\
    \end{tabular}
\caption{Summary of different combinations of the number of variables, $p$, and the number of variable groups, $K$, in $\mSigma$. We set $K$ dynamically in relation to $p$ (e.g. low number of blocks is $3$ for the data with $24$ variables and $12$ for the high-dimensional data with $384$ columns)}
\label{table:exp2-setup}
\end{table}

\begin{figure}[h!]
  \centering
{\includegraphics[width=\linewidth]{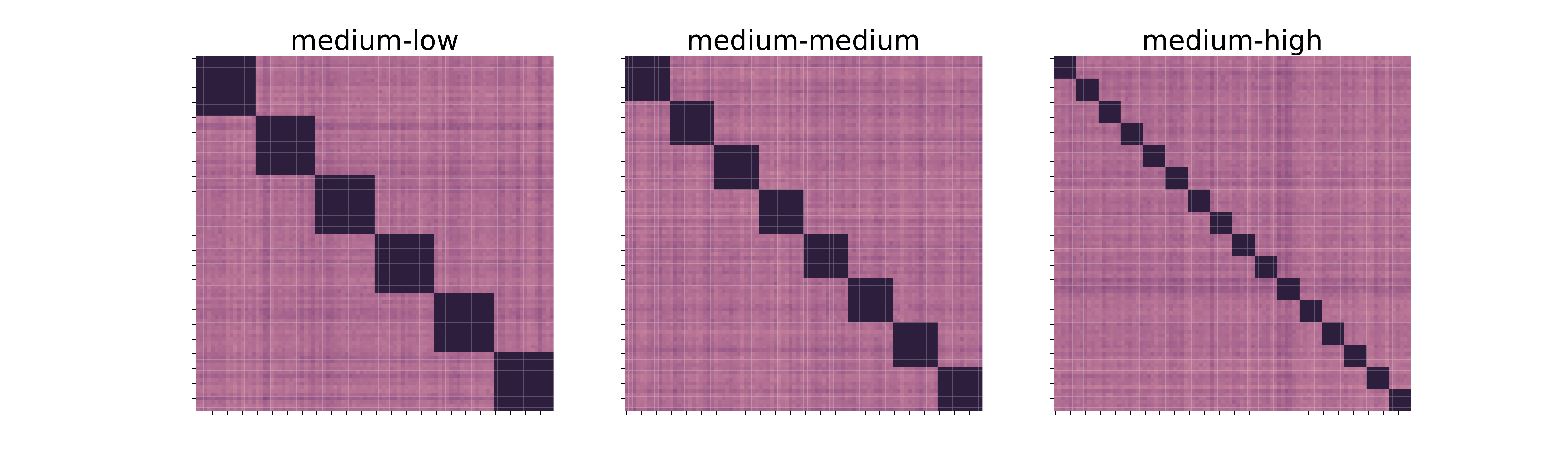}}\vspace{1mm}
  \caption{Examples of the covariance matrices $\mSigma$ used to generate data for the simulation study. These show when the number of columns is set to medium $p=96$  and the number of blocks, $k$ varies from 6 (low), 8 (medium) and 16 (high).  Note the added noise. }
  \label{fig:matrices_2}
\end{figure}

For every combination of $p$ and $K$, we run the analysis $500$ times. In every run, we sample the elements of $\vecmu$ randomly from the uniform $[0, 1]$ distribution. For the block diagonal covariance matrix, each block, $\mSigma_k$, is set to $\mata^\intercal \mata$ where $\mata$ is $p/K$-dimensional square matrix and the entries are generated from uniform $U(1, 2)$. Once each block has been constructed, we add random noise to the overall covariance matrix $\mSigma$ to bring this simulation closer to real-world scenarios. The random noise is equal to $0.5 \mate^\intercal \mate$ where  $\mate$ is $p$-dimensional square matrix and the entries are generated from uniform $U(0, 1)$.

\begin{figure}[h!]
  \centering
  \includegraphics[width=0.48\linewidth]{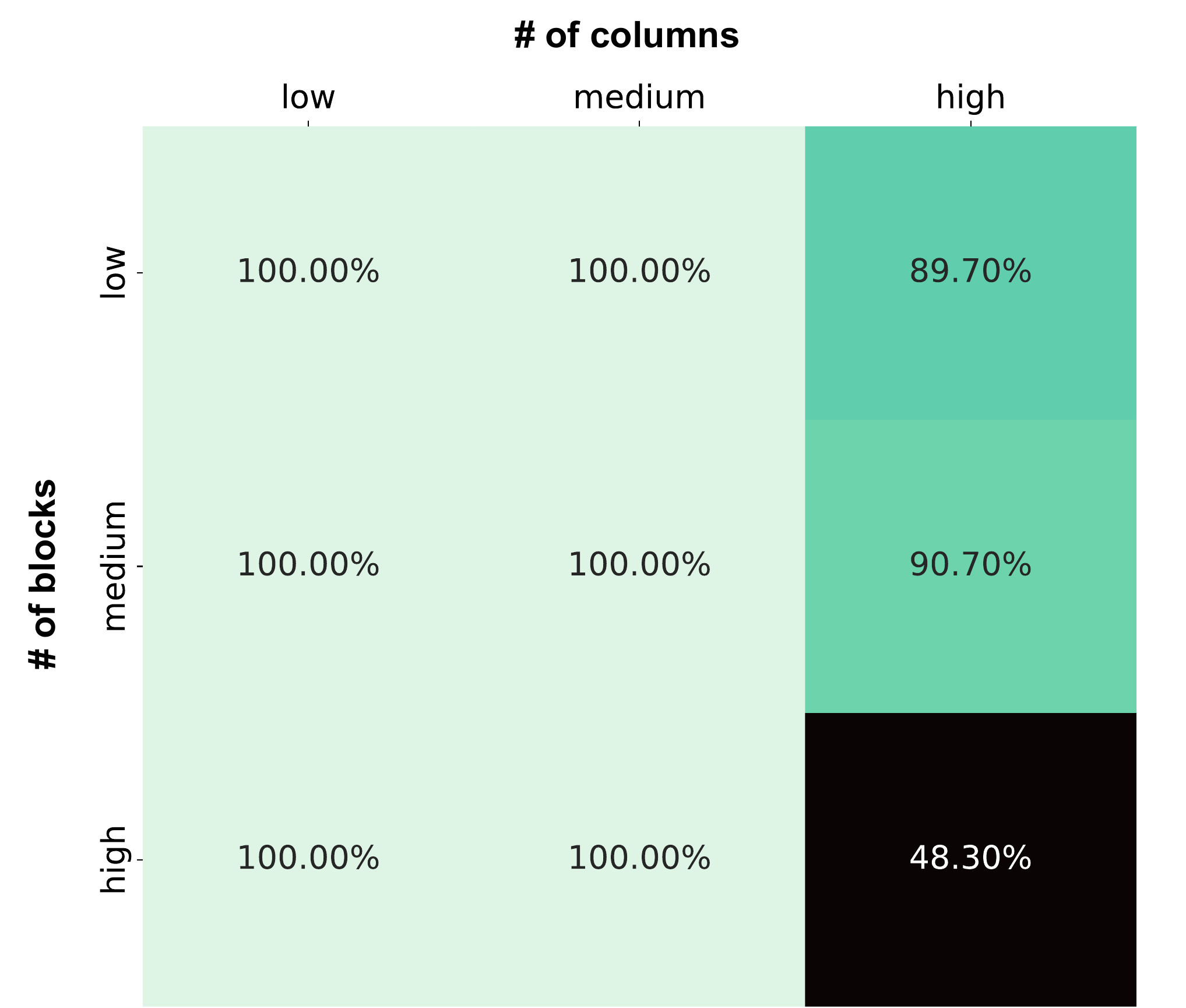} 
\includegraphics[width=0.48\linewidth]{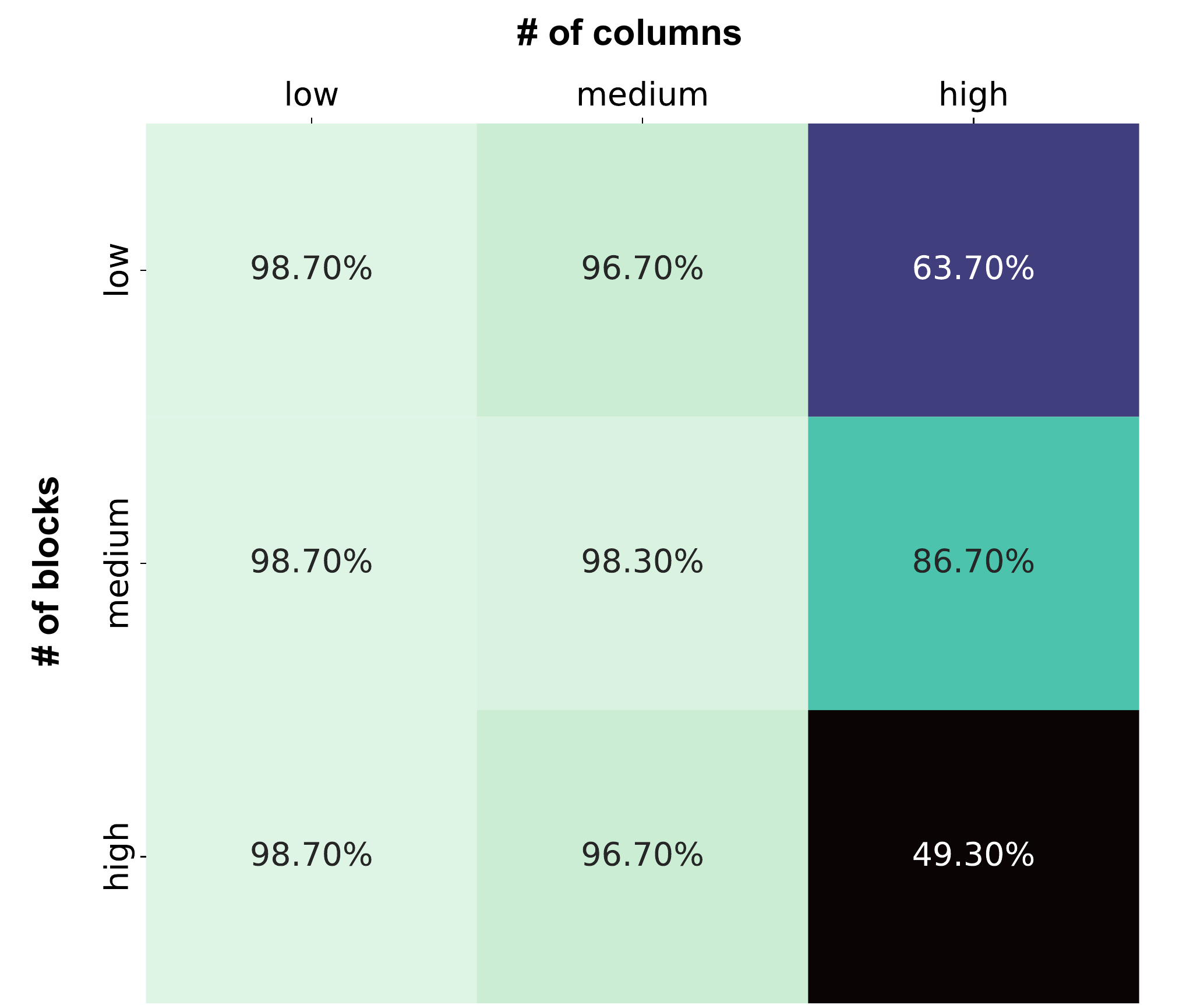}
  \caption{The average accuracy of the hierarchical estimator for the number of column groups $\matd$ while number of variables and the number column groups different  complexities defined by the dimensionality of the dataset(low-high) and the true number of blocks in the covariance matrix $\mSigma$ (number of column clusters). Tested for the scenario of known (\textit{left}) and unknown (\textit{right}) $K$ clusters}
  \label{fig:accuracy_2}
\end{figure}

Figure \ref{fig:accuracy_2} demonstrates the accuracy of the hierarchical estimator for different problem complexities in two scenarios: known vs an unknown actual number of column clusters (blocks of the covariance matrix). In the second case, we pick the number of groups that produces the maximum silhouette score \citep{ROUSSEEUW198753} with the search range $K \pm 2\sqrt{p}$, where $K$ is the number of clusters and $p$ is a dimensionality of the data. 

Figure \ref{fig:accuracy_2} shows that the proposed estimator is accurate for the covariance structures in the low- and medium- setting data but has difficulty in the high-dimensional ($p=384$) setting. Although this might be due the sample size to be fixed at $50$). We see that accuracies for the known vs unknown number of blocks are similar, which signifies the usefulness of the estimator even in the case when the number of column clusters is unknown.

\section{The Bi-Clustering Model with general block diagonal structure } \label{chapter:model}
 
This section extends the proposed block diagonal structure and estimation procedure for model-based bi-clustering. We start by defining a probability density function for a finite $G$-component finite Gaussian mixture model  (GMM),
\begin{equation} \label{gmm}
   g(\vecx_i \mid \vectheta) = \sum_{g = 1}^G \pi_g f\left(\vecx_i \mid \vecmu_g, \mSigma_g \right) 
\end{equation}
where $\pi_g >0$, $\sum_{g=1}^G\pi_g =1$ and $\vectheta$ represents all the model parameters. For bi-clustering model, we assume that $\mSigma_g$ has a block structure as given (\ref{eq:param}). The biggest difference between the proposed model and existing models based on the factor-analyzer model is in how the covariance matrices are constructed.  

There is no closed-form solution for estimating the parameters of the Gaussian mixture model. Parameters of such models are usually estimated by leveraging the expectation-maximization \citep[EM;][]{dempster1977maximum} algorithm, or its variants. The EM algorithm is an iterative algorithm that finds local maximum likelihood estimates of the model parameters $\vectheta$ when the data are incomplete or are treated as incomplete. In the context of Gaussian mixture model, the sample of $N$ observations from (\ref{gmm}) are the observed data. We define the component indicator variable $\mathbf{Z}$ for the row clusters such that $Z_{ig}=1$ if observation $i$ belongs to group $g$ and 0 otherwise. In the clustering context, the group memberships are unknown and thus,
the component indicator variable $Z_{ig}$  are treated as the unobserved latent variable.
% If we have sample of $N$ observations from (\ref{gmm}). We apply the EM by letting $z_{ig}$  for $i=1, \dots, N$ and $g=1, \dots, G$ be the latent variable with $z_{ig}$ determining the probability of $\vecx_i$ to be generated from $g$-th component. 

The EM algorithm alternates between the expectation (E-step) and maximization (M-step) steps. In the E-step, we compute the expected value of the complete data log-likelihood function using the current parameter values $\vectheta^{(t)}$. In the M-step, the algorithm finds the estimate of $\vectheta$ that maximizes the expected value of the log-likelihood from the E-step. For a $G$-component GMM, the expected value of the complete data log-likelihood function is
\begin{equation*} \label{equation:expectation}
     \sum_{i=1}^N  \sum_{g=1}^G  \hat{z}_{ig} [\log \pi_g - \frac{1}{2} \log |\mSigma_g| - \frac{1}{2}\tr\left[\mSigma_g^{-1} (\vecx_i - \vecmu_g) (\vecx_i - \vecmu_g)^\intercal \right],
\end{equation*}
where expected values of the latent variable $z_{is}$ are
\begin{equation}     \label{z}
    \hat{z}_{ig} = P( Z_{ig} = 1| \vecx_i, \vectheta ) = \frac{\pi_g f( \vecx_i|\vecmu_g, \mSigma_g)}{\sum_{h=1}^G \pi_h f( \vecx_i \mid \vecmu_h, \mSigma_h)}.
\end{equation}
%\begin{equation}
%    Q(\theta|\theta^t ) = E_{Z|X, \theta^\intercal}[\log L(\theta;X, Z)] =  \sum_{i=1}^N  E_{Z_i|X_i, \theta^t }[\log L(\theta;X_i, Z_i)] = 
%   \end{equation}
%\begin{equation*}
%     \sum_{i=1}^N  \sum_{s=1}^S P(z_{is} = 1|X_i, \vectheta ) [\log \pi_s - \frac{1}{2} \log |\mSigma_g| - \frac{1}{2}\tr\left[\mSigma_g^{-1} (\vecx_i - \mu_g) (\vecx_i - \mu_g)^\intercal \right].
%\end{equation*}
%\begin{equation*}
%     \sum_{i=1}^N  \sum_{s=1}^S  z_{ig} [\log \pi_s - \frac{1}{2} \log |\mSigma_g| - \frac{1}{2}\tr\left[\mSigma_g^{-1} (\vecx_i - \mu_g) (\vecx_i - \mu_g)^\intercal \right].
%\end{equation*}

For the M-step, the updates for the $\vectheta$ are:
\begin{equation}
   \hat{\pi}_g = \frac{1}{N} \sum_{i=1}^N \hat{z}_{ig}, \quad \hat{\vecmu}_g = \frac{ \sum_{i=1}^N \hat{z}_{ig} \vecx_i}{\sum_{i=1}^N \hat{z}_{ig}} .
\end{equation}
The maximum likelihood estimate of unconstrained $\mSigma_g$ is given by:
\begin{equation}
   \hat{\mSigma}_g= \mats_g = \frac{ \sum_{i=1}^N \hat{z}_{ig}(\vecx_i - \vecmu_g)(\vecx_i - \vecmu_g)^\intercal }{ \sum_{i=1}^N \hat{z}_{ig} } .
    \label{em_sigma}
\end{equation}
To estimate a block-diagonal  $\mSigma_g$, we input the $\mats_g$ to one of the three proposed column clustering estimators (e.g. greedy, hierarchical, numerical). Then we compute the estimate of the block covariance matrix for the $g^{th}$ component with $K$ blocks as:
\begin{equation}
\sum_{k=1}^K \matd_{kg} \mats_g \matd_{kg}.
\end{equation}
%In the case of the common covariance matrix, we compute $\mSigma = \sum_{g=1}^G\pi_g\Sigma_G$ and repeat the steps outlined above.  \textcolor{blue}{ do use common covariance matrix? }

For initialization, we use the K-means algorithm \citep{forgy1965cluster} to provide initial assignment of the latent variables $z_{ig}$. We use Aiken's acceleration based criteria \citep{bohning94} to determine convergence of the EM algorithm  using a tolerance of $10^{-4}$. To obtain row clustering memberships, we take the argmax over the set $\{ \hat{z}_{i1}, \ldots, \hat{z}_{iG} \}$ and the column groups are obtain by examining at the matrices $\matd_{k}$.

\subsection{Estimating the Number of Row Clusters}
\label{section:optimal-clusters-sim}
%======================================================================

Here, we study the estimation of the number of clusters by the proposed  model and it's accuracy of selecting the number of row clusters given data generated from a GMM with $G=3$ components. We consider two scenarios where generate data from  $\mathcal{N}(\vecmu_g, \mSigma_g)$ where the covariance matrix has a block diagonal structure and a general structure (or non-block diagonal structure).

We choose similar parameter values to the simulation from \citep{Tu2022AFO}. In the first scenario, we set
\begin{equation*} \begin{split}
 \vecmu_1 &= \left(-5,-4,-3,-2,-1,0,1,2 \right)^\intercal, \\
 \vecmu_2 &=\left(0,1,2,3,4,5,6,7\right)^\intercal, \\
 \vecmu_3 &= \left(5,6,7,8,9,10,11,12 \right)^\intercal
\end{split} \end{equation*}
and the covariance matrices are block diagonal. 
In particular we set $\mSigma_1= \diag( \mSigma_{11} ,  \mSigma_{12} ,  \mSigma_{13} )$  where 
\begin{equation*} 
 \mSigma_{11} 
 = \left[ \begin{array}{ccc}
2.5 & 0.5  & 0.5   \\
0.5 &  3.5  & 0.5  \\
0.5 &  0.5 &  4.5
\end{array} \right],
\quad
 \mSigma_{12} 
 = \left[ \begin{array}{ccc}
2 & 1  & 1   \\
1 &  2  & 1  \\
1 &  1 &  2
\end{array} \right],
\quad
 \mSigma_{13} 
 = \left[ \begin{array}{cc}
3.5 & 3.0    \\
3.0 &  3.9   \\
\end{array} \right],
\end{equation*}
and set $\mSigma_2= \diag( \mSigma_{21},  \mSigma_{22},  \mSigma_{23} )$ where
\begin{equation*} 
\mSigma_{21} 
 = \left[ \begin{array}{ccc}
4.5 & -2  & 1   \\
-2 &  4.5  & 2  \\
 1 &  2 &  4.5
\end{array} \right],
\quad
 \mSigma_{23} 
 = \left[ \begin{array}{cc}
4 & 2    \\
2 &  4    \\
\end{array} \right],
\quad
 \mSigma_{22} 
 = \left[ \begin{array}{ccc}
4 & 3  & 3   \\
3 &  3.5  & 3  \\
3 &  3 &  4
\end{array} \right],
\end{equation*}
and set $\mSigma_3= \diag( \mSigma_{31},  \mSigma_{32})$ where
\begin{equation*} 
\mSigma_{31} 
 = \left[ \begin{array}{ccccc}
2.1 & 2  & 2 & 2 & 2  \\
 2 &  2.5  & 2 & 2 & 2  \\
 2 &  2 &  3 & 2 & 2  \\
 2 &  2 &  2 & 5 & 2   \\
 2 &  2 &  2 & 2 & 4
\end{array} \right],
\quad
 \mSigma_{32} 
 = \left[ \begin{array}{ccc}
2 & 1  & 1   \\
1 &  3.5  & 1  \\
1 &  1 &  2.5
\end{array} \right].
\end{equation*}

In the second scenario, we set $\mSigma_g$ to be different random positive-definite matrices generated by setting $\mSigma_g = \mata_g^T\mata_g$ where the elements of $\mata_g$ are generated independently  from the uniform distribution on the interval $(0.0, 1.0)$. This comparison considers what happens when the data is generated from a model that violates the bi-clustering framework or assumption of block-diagonal covariance matrix.

In both scenarios, we set $\pi_1=\pi_2=\pi_3=1/3$ and we vary the number of observations per component. Then we chose the number of row cluster by iterating over the search space $\hat{G} \in \left\{1\dots 5\right\}$ and applying the BIC \citep{bic, schwartz78} criterion to select the number of components. For both scenarios we replicate the experiment $100$ times. We compare performance of the proposed method to the performance of a GMM.

\begin{figure}[h!]
\centering
\includegraphics[width=0.48\linewidth]{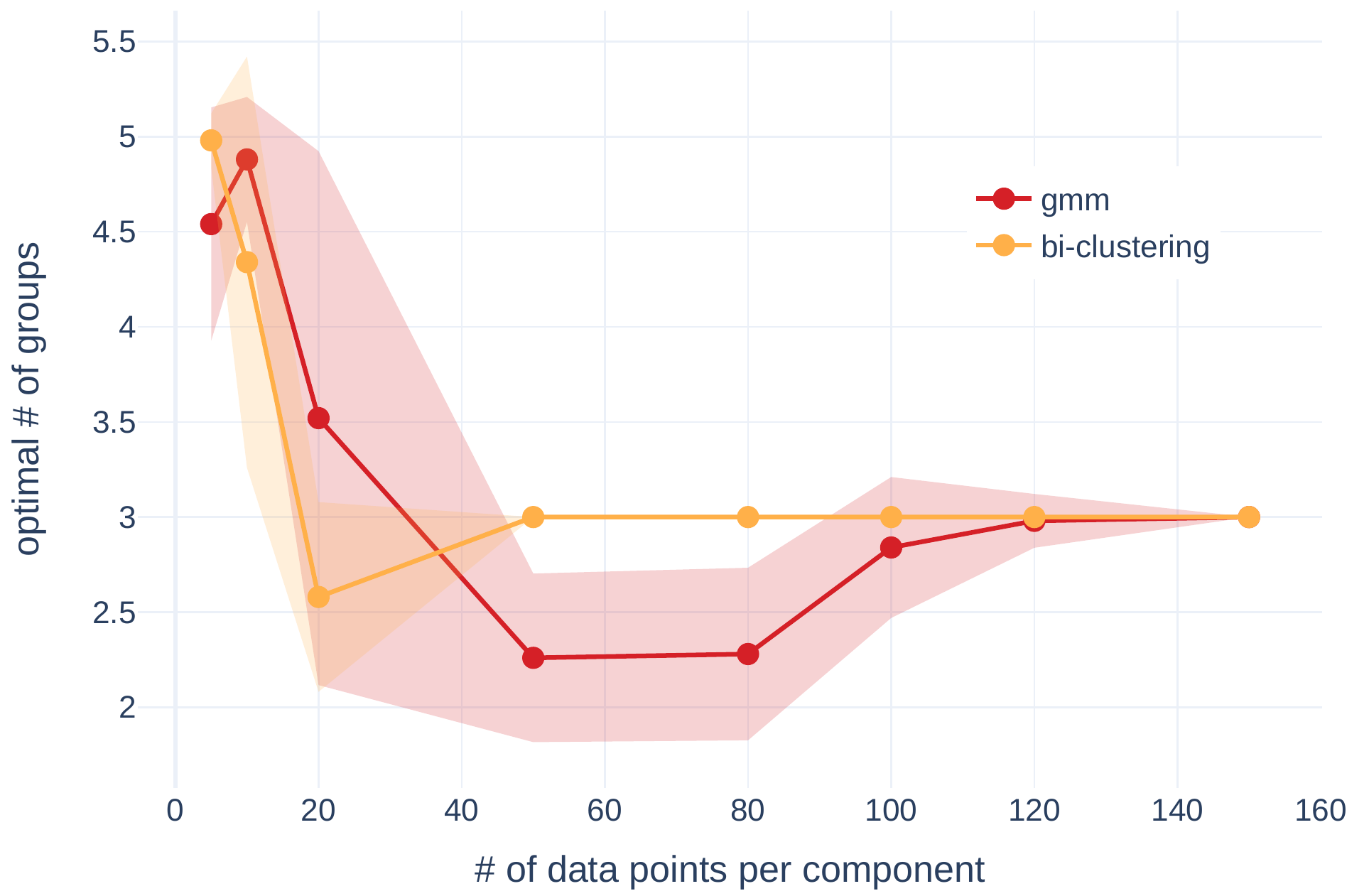} 
\includegraphics[width=0.4\linewidth]{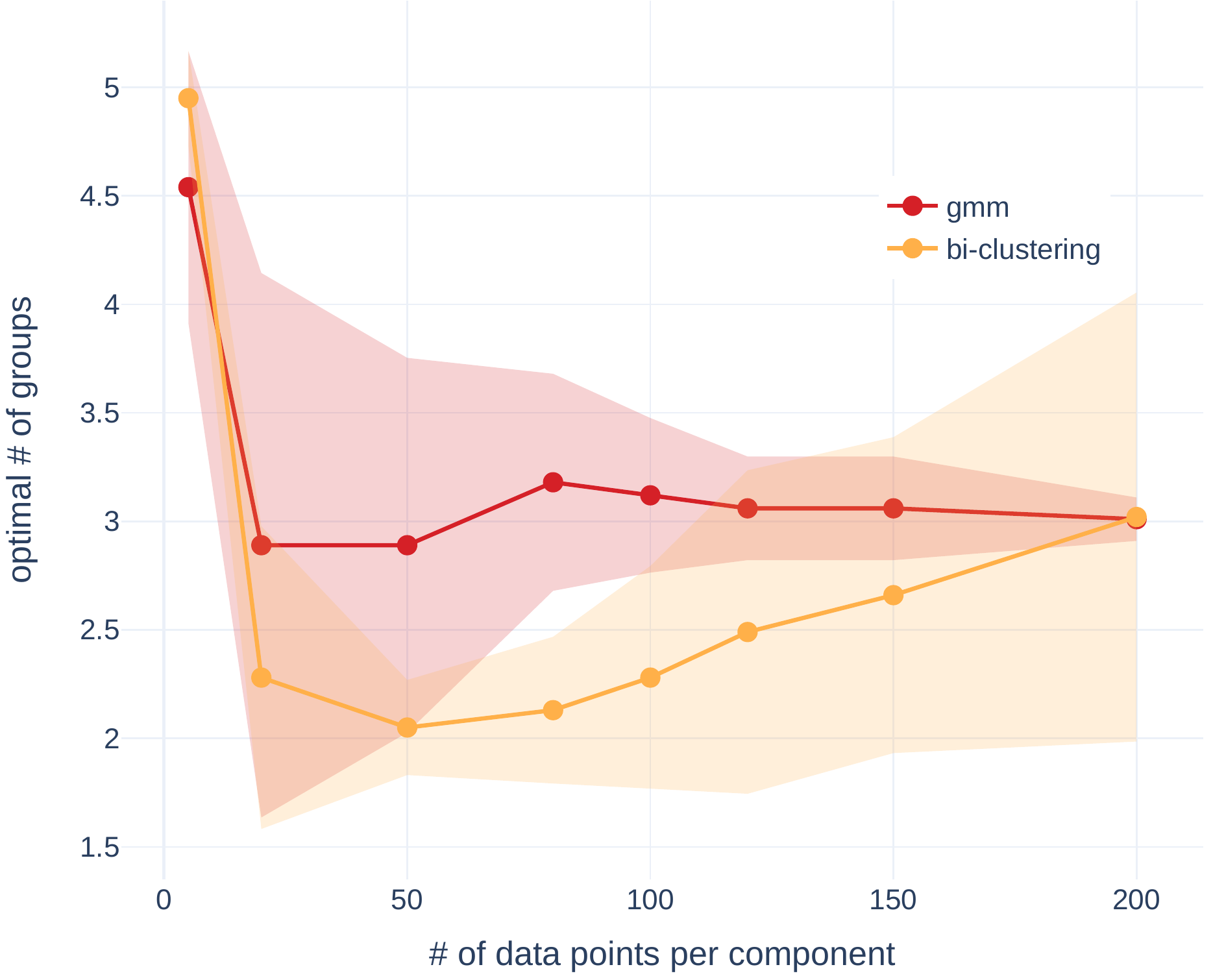} 
% \captionsetup{justification=centering,margin=2cm}
\caption{Recovery of the number of groups when the covariance structure is block diagonal (\textit{Left}) and a random positive definite matrix (\textit{Right}). Both panels display the average and plus/minus the standard deviation for a GMM and the proposed bi-clustering model when $G=3$ while varying the number of observations within each component.
}
\label{fig:num-clusters-recovery}
\end{figure}

Figure \ref{fig:num-clusters-recovery} shows the average estimated number of components with band showing plus/minus the standard deviation for bi-clustering and GMM under the two scenarios. Both panels show that the models select the true number of row clusters more often as the size of input data grows. However, in the left panel the bi-clustering model arrives to $100$\% accuracy earlier. On the other hand, the right panel shows that a general covariance structure causes issues for the bi-clustering model which needs lots variable groups to achieve a good fit of the data. This provides some evidence that in the presence of a block-diagonal covariance matrix, the bi-clustering is efficient in determining the number of row clusters. In addition, Figure \ref{fig:num-clusters-acc} in \ref{accuracy-plots} summarizes the accuracy of identifying the correct number of groups.

\subsection{Bi-Clustering Applications}
\label{section:clustering}

%Here, we aim to conduct experiments to evaluate how accurate and efficient the proposed approach is in the clustering scenario. In the \ref{section:clustering}, we compare the performance of our algorithm to the performance of two clustering algorithms: $K$-means \citep{forgy1965cluster} and GMM \citep{reynolds2009gaussian}; and two bi-clustering approaches: spectral co-clustering \citep{Dhillon2001} and factor-analyzer based approach  by \citep{Tu2022AFO}. We assess the accuracy by using the percentage (\% acc) of observations that corresponds the true component. For every generated dataset, we find that our algorithm is comparable or better in accuracy. 

In this subsection, we demonstrate the applicability of the proposed biclustering algorithms on benchmark datasets as well as high dimensional datasets from bioinformatics and topics modelling. We examine the row clustering performance of the proposed bi-clustering approach i.e., the Gaussian mixture model with a block-diagonal covariance matrix estimated by the hierarchical method. We compare the results of the proposed approach with $K$-means \citep{forgy1965cluster}, the unrestricted Gaussian mixture model \citep{reynolds2009gaussian}, and two other bi-clustering methods (UCUU by \cite{Tu2022AFO} and spectral co-clustering \cite{Dhillon2001}). We use $K$-means, GMM and spectral co-clustering implementations with their default parameters from the \texttt{sklearn library} \citep{scikit-learn}. Additionally, we implemented the factor-analyzer UCUU model by \cite{Tu2022AFO} in Python. 

\subsubsection{Bi-clustering benchmark datasets}
Here, we applied the algorithms on well-known low-dimensional benchmark datasets: \texttt{Wine} \citep{cortez2009modeling}, \texttt{Olive} \citep{forina1982pattern} and \texttt{Ecoli} \citep{nakai1991uci}. We scale the data to have the mean equal to $0$ and variance equal to $1$ prior to clustering. We compare the clustering performanace using the adjusted rand index (ARI) \citep{rand71, hubert85, zhang2012generalized}, accuracy and computational time. The results are summarized in Table \ref{table:datasets-metrics}.

\begin{table}[h!]
    \scriptsize
    \centering
        \caption{Average adjusted rand index (ARI), accuracy and execution time over 10 different runs of all algorithms on the given datasets.}
    \label{table:datasets-metrics}
    \scalebox{0.95}{
    \begin{tabular}{ c | c c c | c c c | ccc }
        Datasets 
            &\multicolumn{3}{c}{Wine}
            &\multicolumn{3}{c}{Olive}
            &\multicolumn{3}{c}{Ecoli}
 \\
        %\midrule
        & ARI & \%, acc & time
        & ARI & \%, acc & time 
        & ARI & \%, acc & time  \\ 
                & & & (in sec)
        & &  & (in sec)
        &  &  & (in sec) \\ \hline
        \hline
        K-means
        & $0.897$ & $96.6 \%$ & $0.015$
        & $0.448$ & $76.5\%$ & $0.047$
        & $0.509$ & $65.2\%$ & $0.036$ \\
        GMM
        & $0.831$ & $92.7 \%$ & $\mathbf{0.009}$
        & $\mathbf{0.601}$ & $78.1\%$ & $\mathbf{0.019}$
        & $0.646$ & $74.4\%$ & $\textbf{0.015}$ \\
        UCUU Model
        & $\mathbf{0.948}$ & $\mathbf{98.3} \%$ & $4.320$
        & $0.517$ & $79.1\%$ & $19.62$
        & $-$ & $-$ & $-$\\
        Proposed
        & $0.945$ & $98.3\%$ & $0.120$
        & $0.574$ & $\mathbf{80.4\%}$ & $0.169$
        & $\mathbf{0.656}$ & $\mathbf{76.2}\%$ & $0.138$\\
        Spectral co-clustering
        & $0.738$ & $90.9\%$ & $0.020$
        & $0.237$ & $57.3\%$ & $0.042$
        & $0.394$ & $56.4\%$ & $0.048$ \\
    \end{tabular}
    }
    \vspace{0.2cm}

\end{table}

The proposed algorithm has similar accuracy compared to known row clustering algorithms and other bi-clustering approaches
and even outperforms others on the \texttt{Ecoli} dataset.
However, is almost $10\times$ slower in terms of execution time than other listed approaches except for the UCUU model. The UCUU model, which is most similar algorithm to our proposed method, runs $400\times$ slower than $K$-means, GMM and spectral co-clustering. Thus, the proposed approach has  an advantage in terms of lower execution time compared to the bi-clustering UCUU model, which is very similar in utility.

\subsubsection{Bi-clustering in Bioinformatics}
Biclustering is used in bioinformatics to simultenously cluster observations and genes. Identifying groups of highly correlated genes that have different correlation structure in different groups of individuals (i.e., disease vs healthy individuals or between individuals with subtypes of diseases) can shed light into underlying biological mechanisms of disease development.

Here, we compare the algorithms' performance on the high-dimensional genomics classification problems using two benchmark gene expression datasets \texttt{Alon} \citep{alon} and \texttt{Golub} \citep{golub99} available in the {\sf R} package \texttt{plsgenomics}\citep{plsgenomics}. We applied our approach on two versions of the datasets: (i) a subset of 100 genes selected using ANOVA F-test \citep{anova} which from here on is referred to as \texttt{Alon}$_{100}$ and (ii) full datasets where all available genes are used. The ANOVA F-test selects the set of $K = 100$ features with highest ratio of explained to unexplained variance between the given ground truth groups.
We applied our approach other competing approaches. Unfortunately, we were unable to run the UCUU model by \cite{Tu2022AFO} on the full set of features because of $(i)$ increasing computational time; $(ii)$ authors' note that this algorithm is not designed for the scenario when the number of features is significantly higher than the number of observations. Therefore, UCUU model was only applied to \texttt{Alon}$_{100}$.

% The preprocessed version of \texttt{Alon} data from \cite{ McNicholas2008ParsimoniousGM} consists of 42 tumorous and 22 normal observations of 461 genes. \textcolor{blue}{***Is the results presented here using full 2000 from plsgenomics or 461 from the old version of the dataset that I received from Paul?***} 

  The performance of the proposed algorithm along with comparisons with other approaches in terms of ARI and \% accuracy on Alon$_{100}$ and Alon$_\text{full}$ datasets are provided in Table \ref{table:high-dimensional metrics}.

% In the second experiment, we compare the performance of the same pool of algorithms on a more complex task of high-dimensional data clustering. 
\begin{table}[h!]
 \scriptsize
    \centering
        \caption{ARI and accuracy of different algorithms on given high-dimensional datasets and their preprocessed versions with $100/150$ features selected with ANOVA F-test\cite{anova}. }    \label{table:high-dimensional metrics}
    \begin{tabular}{ c | cc | cc | cc | cc}
        Datasets   
            &\multicolumn{2}{c}{Alon$_{100}$}
            &\multicolumn{2}{c}{Golub$_{150}$}
            &\multicolumn{2}{c}{Alon$_\text{full}$}
            &\multicolumn{2}{c}{Golub$_\text{full}$} \\ 
        %\midrule
        & ARI & \%, acc 
        & ARI & \%, acc 
        & ARI & \%, acc 
        & ARI & \%, acc \\ \hline
        K-means
        & $0.510$ & $86.0\%$
        & $0.892$ & $97.3\%$
        & $-0.002$ & $55.3\%$
           & $0.162$ & $70.8\%$\\
        GMM
        & $0.545$ & $87.0\%$
        & $\mathbf{0.935}$ & $\mathbf{98.4}\%$ 
        & $0.001$ & $56.4\%$ 
        & $0.161$ & $53.2\%$\\
         UCUU Model
        & $\mathbf{0.609}$ & $\mathbf{89.2}\%$
        & $0.892$ & $97.3\%$ 
        & $-$ & $-$
        & $-$ & $-$\\
        Proposed
        & $0.541$ & $87.0\%$
        & $0.892$ & $97.3\%$
        & $-0.006$ & $54.8\%$
        & $\mathbf{0.212}$ & $\mathbf{73.6\%}$\\
        Spectral co-clustering
         & $0.541$ & $87.0\%$
        & $0.892$ & $97.3\%$ 
        & $\mathbf{0.040}$ & $\mathbf{62.9\%}$
         & $0.185$ & $72.2\%$\\
    \end{tabular}
\end{table}

In the case of \texttt{Alon}$_{100}$, the UCUU model had a slightly higher ARI compared to the proposed model. However, our proposed method takes substantially less time to fit. Unfortunately, the performance of all methods degraded significantly on the full \texttt{Alon} dataset, with the Spector co-clustering method achieving maximum accuracy of $62.9$\%. 

\begin{figure}
\begin{subfigure}{0.5\textwidth}
\includegraphics[scale=0.48]{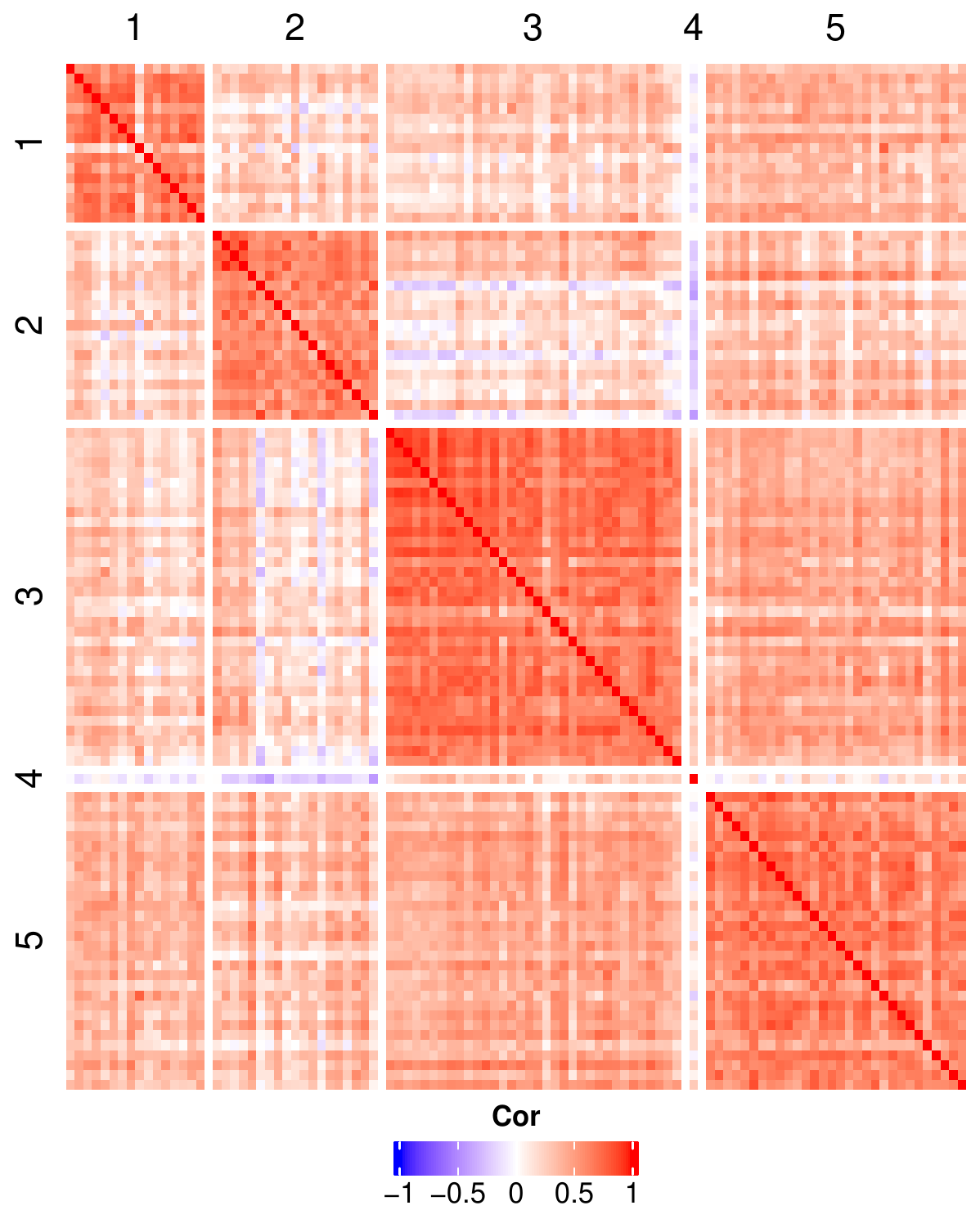}
\caption{Observed correlation of predicted Group 1}
\end{subfigure}
\begin{subfigure}{0.5\textwidth}
\includegraphics[scale=0.48]{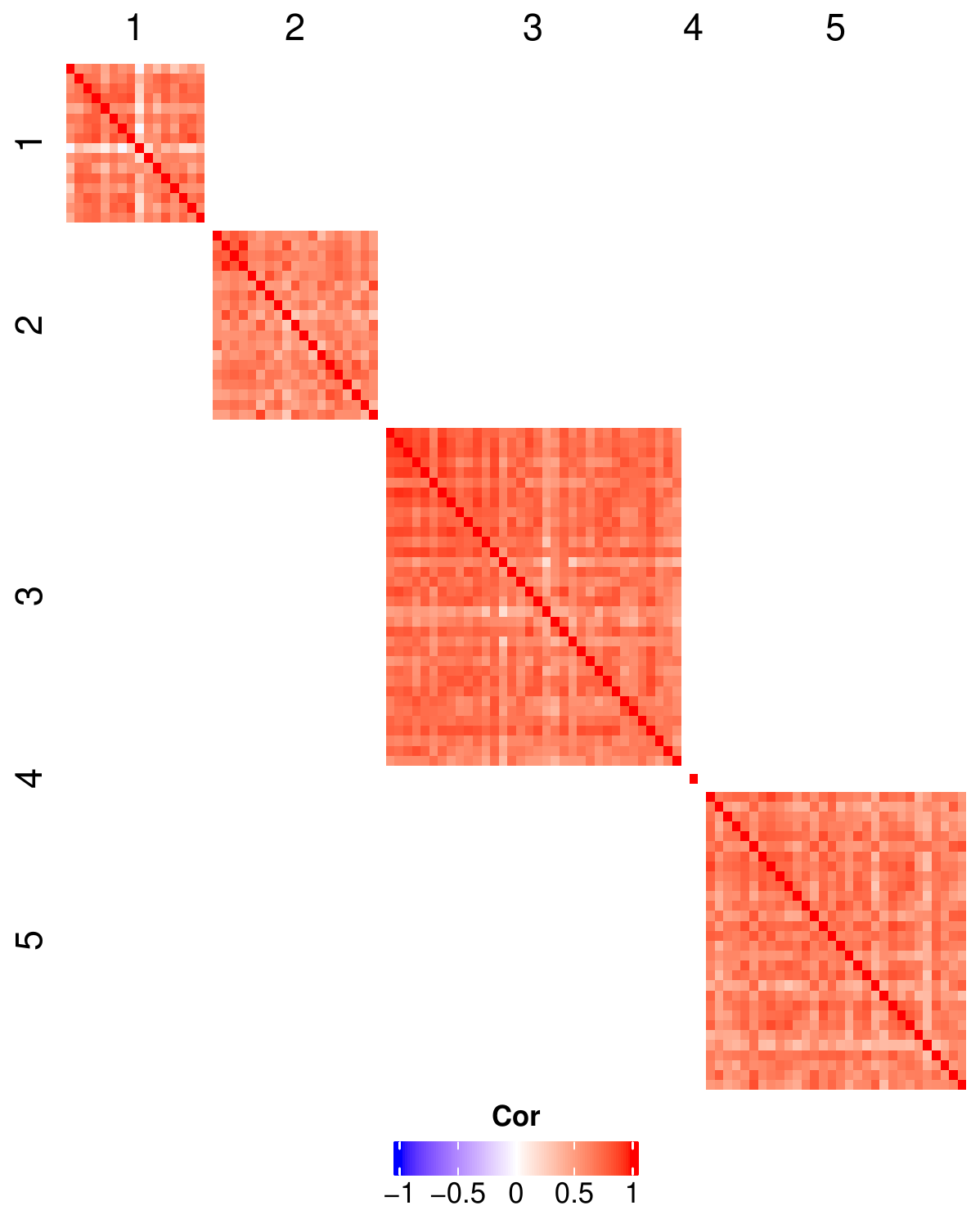}
\caption{Estimated correlation of predicted Group 1}
\end{subfigure}
\begin{subfigure}{0.5\textwidth}
\includegraphics[scale=0.48]{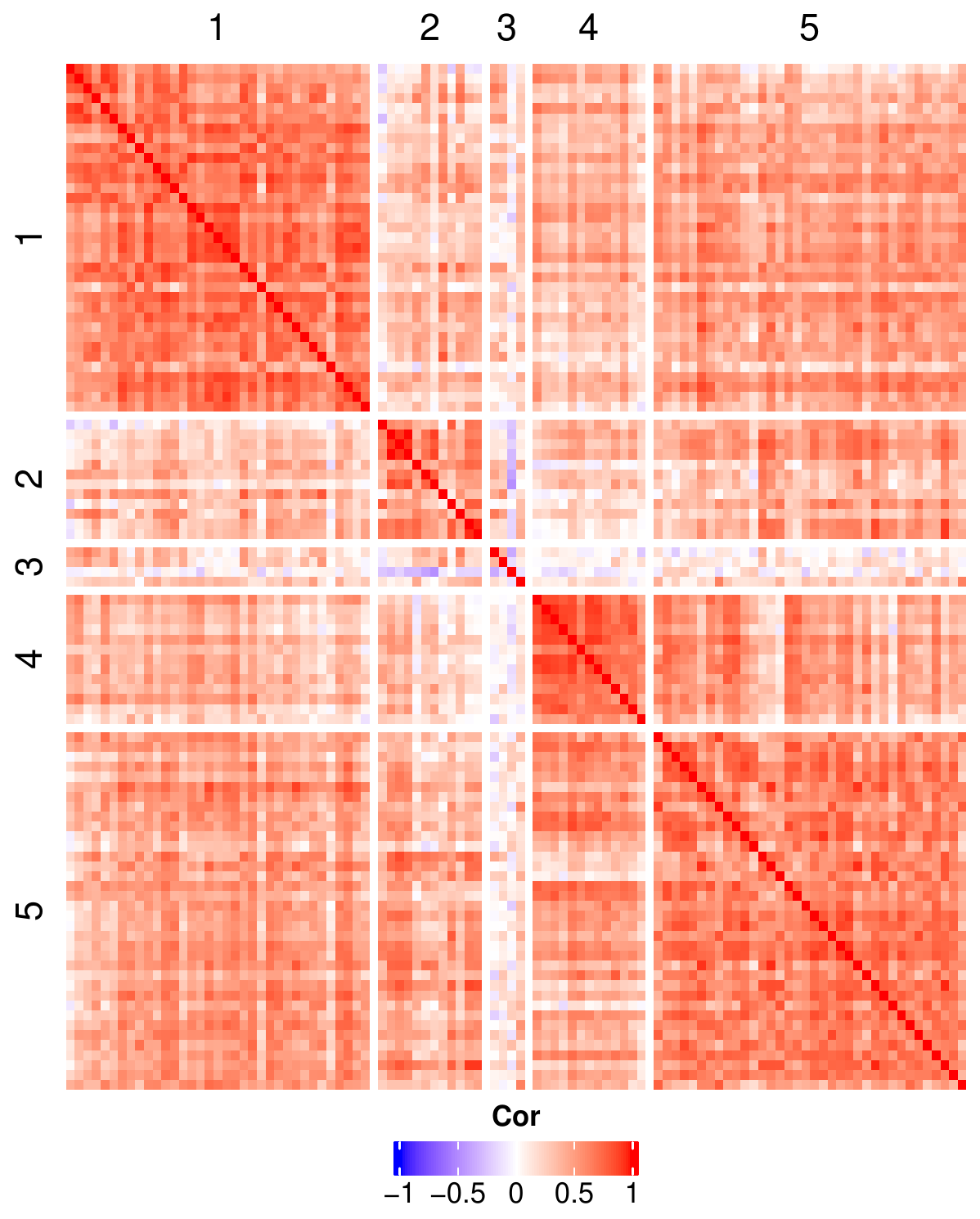}
\caption{Observed correlation of predicted Group 2}
\end{subfigure}
\begin{subfigure}{0.5\textwidth}
\includegraphics[scale=0.48]{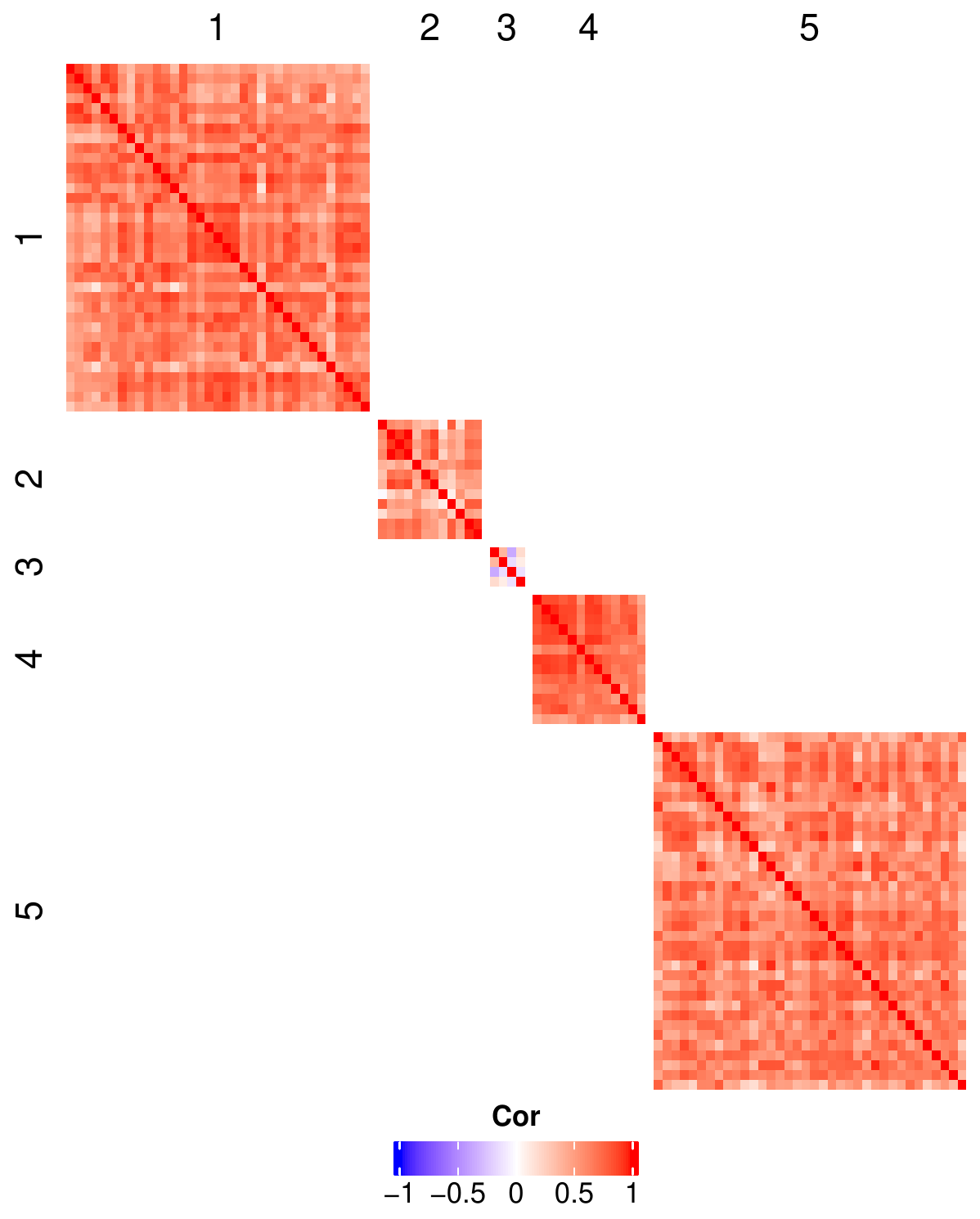}
\caption{Estimated correlation of predicted Group 2}
\end{subfigure}
\caption{True and estimated correlation matrices of the predicted groups from the \texttt{Alon}$_{100}$ dataset. Note that most tumour samples were classified as Group 1 and most healthy samples were classified as Group 2.}\label{alon}
\end{figure}

In Figure \ref{alon}, we visualize the true and estimated correlation structures in the two estimated clusters of the \texttt{Alon}$_{100}$ dataset. Note that column permutations are done to achieve a checkerboard-like structure as illustrated. As can be seen in Figure \ref{alon}, the estimated column clusters are grouping variables that have high correlation (either positive or negative). This can provide valuable biological insight. For example, the variables \texttt{X53586} (mRNA for integrin alpha 6), \texttt{X54942} (ckshs2 - mRNA for Cks1 protein homologue) and \texttt{X54941} (ckshs1 - mRNA for Cks1 protein homologue) are all assigned to column cluster 5 in Group 1 (with majority of tumour samples) indicating that they have a high correlation in tumour samples. Both integrin alpha 6 \citep{beaulieu2020} and Cks1 protein \citep{shapira2005} have previously been identified to play roles in colorectal cancer. On the other hand, in Group 2 (with majority of healthy samples), these variables were assigned to different column clusters: \texttt{X53586} was assigned to column cluster 1, \texttt{X54942} was assigned to column cluster 5 and \texttt{X54941} was assiged to column cluster 3, indicating that these variables have little to no correlation in healthy samples.

\texttt{Golub} dataset \citep{golub99} comprises 2030 gene expressions of 72 patients, 47 of them are diagnosed with acute lymphoblastic leukemia (ALL) and 25 with acute myeloid leukemia (AML). Similar to the \texttt{Alon} dataset, we construct \texttt{Golub}$_{150}$ with subset of 150 features selected with ANOVA F-test \citep{anova}. We compare the performance of the proposed approach with competing approach on both the \texttt{Golub}$_{150}$ and the full dataset. Here, for the \texttt{Golub}$_{150}$ dataset, the GMM has the largest ARI slightly higher compared to our proposed model. However, due to the block diagonal covariance structure, our proposed method is more parsimonious than GMM. Moreover, the column clusters can provide valuable insights as illustrated below which the GMM lacks. Similar to the full \texttt{Alon} data, the performance of all methods degraded significantly on the full \texttt{Golub} dataset - in this case substantially for the GMM, while our proposed method achieves maximum accuracy of $73.6$\% among the competiting models.

% The results here are similar to those from the \texttt{Alon} dataset: five methods demonstrate the identical performance of Golub$_{150}$ dataset, and the performance of all the methods degrades on the full dataset. Here, we observe that  both bi-clustering algorithms (proposed and Spectral Co-Clustering) perform better than row clustering ones. \textcolor{blue}{**Could we confirm this is the same for Alon dataset as well now that we are working with 2000 genes for alon too*** This can be attributed to the bi-clustering ability to homogenize high-dimensional and heterogeneous data.}

\begin{figure}
\begin{subfigure}{0.5\textwidth}
\includegraphics[scale=0.5]{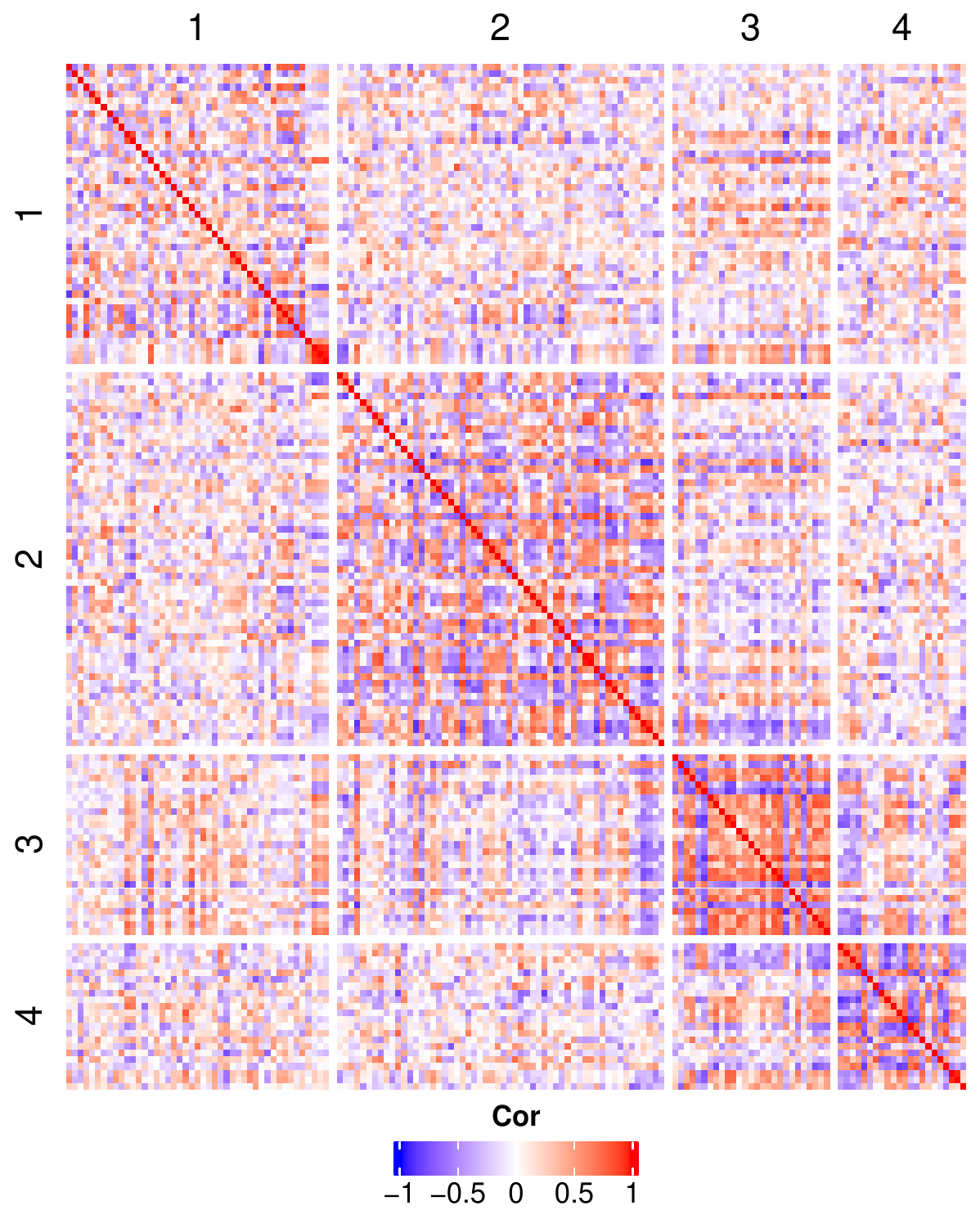}
\caption{Observed correlation of predicted Group 1}
\end{subfigure}
\begin{subfigure}{0.5\textwidth}
\includegraphics[scale=0.5]{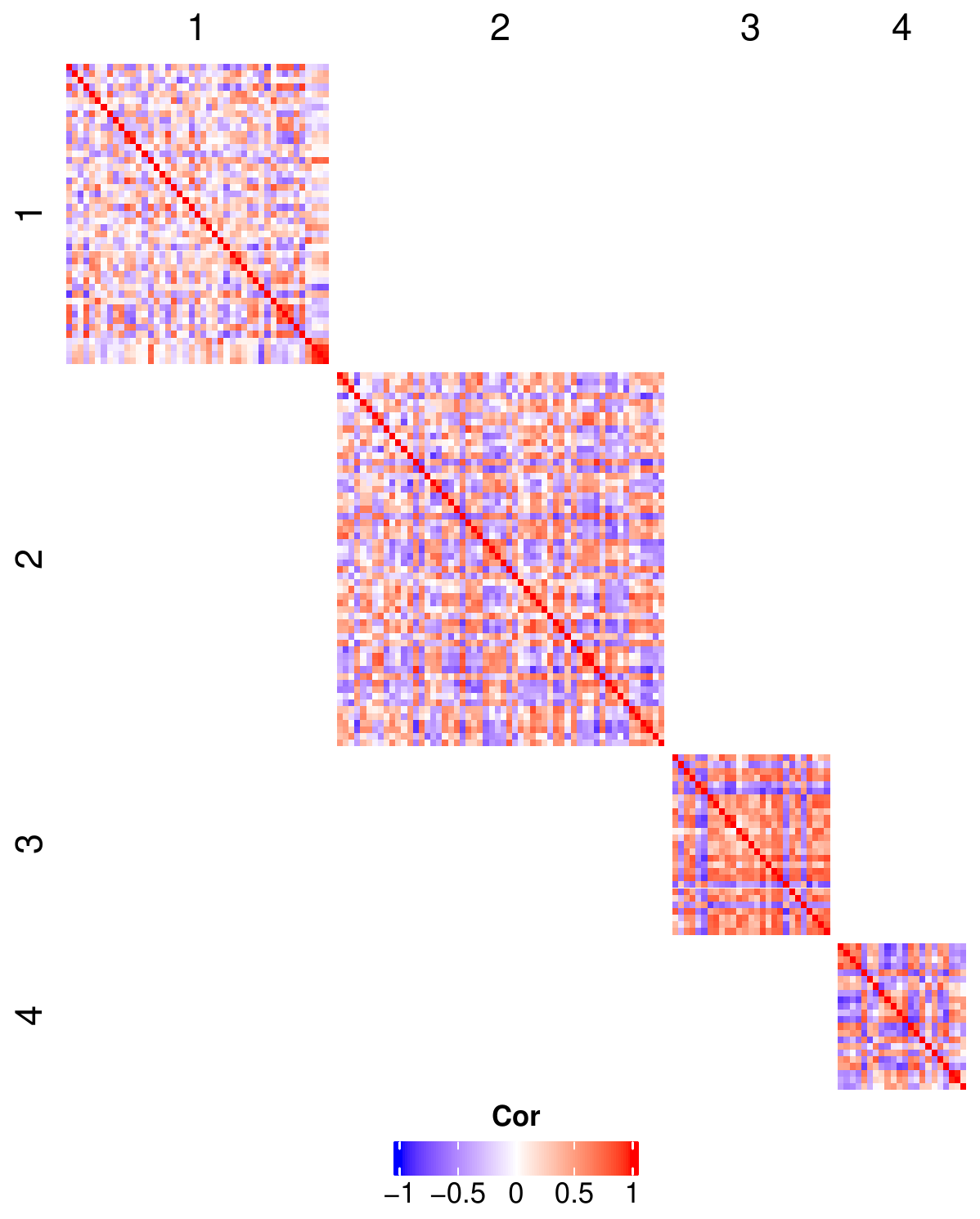}
\caption{Estimated correlation of predicted Group 1}
\end{subfigure}
\begin{subfigure}{0.5\textwidth}
\includegraphics[scale=0.5]{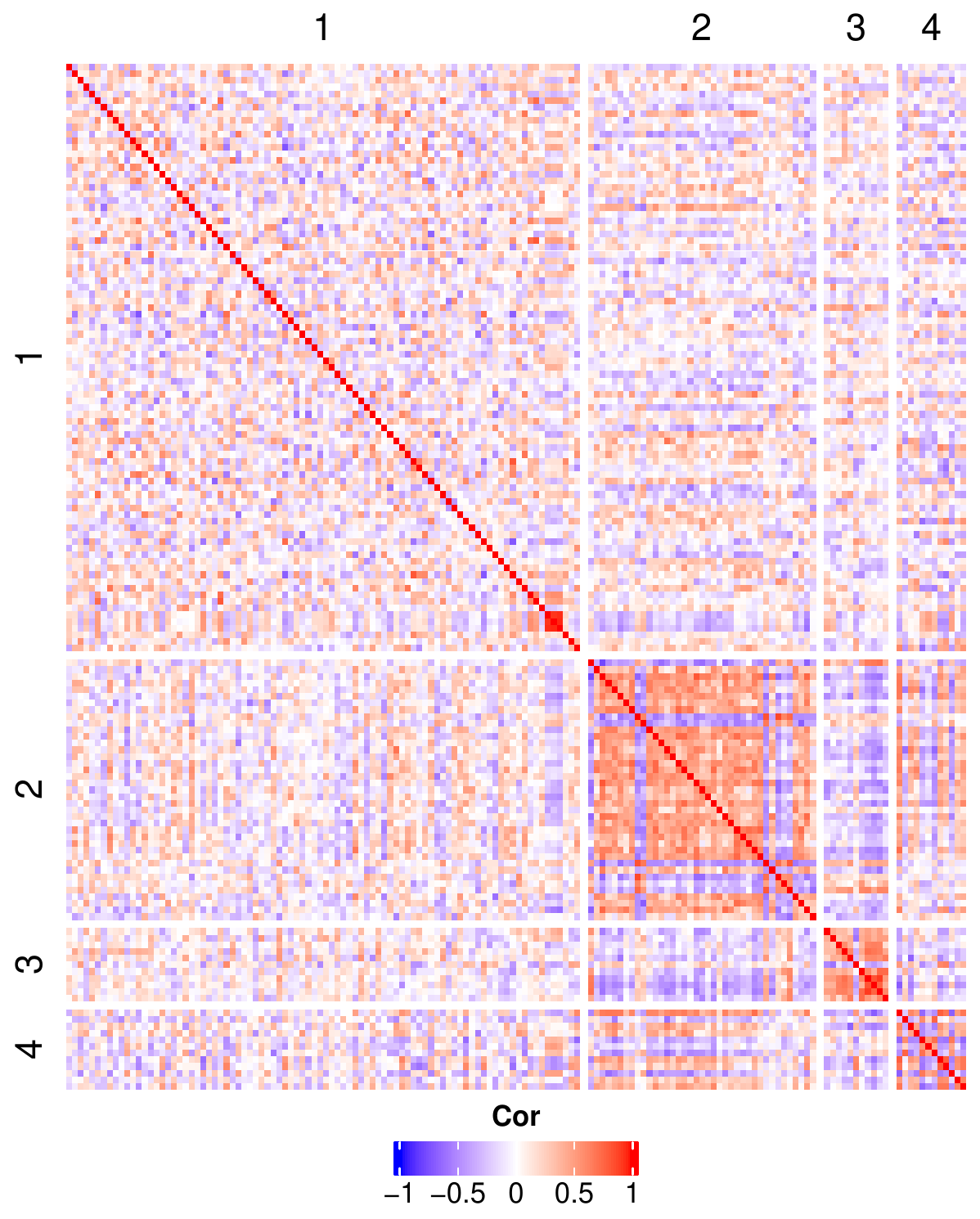}
\caption{Observed correlation of predicted Group 2}
\end{subfigure}
\begin{subfigure}{0.5\textwidth}
\includegraphics[scale=0.5]{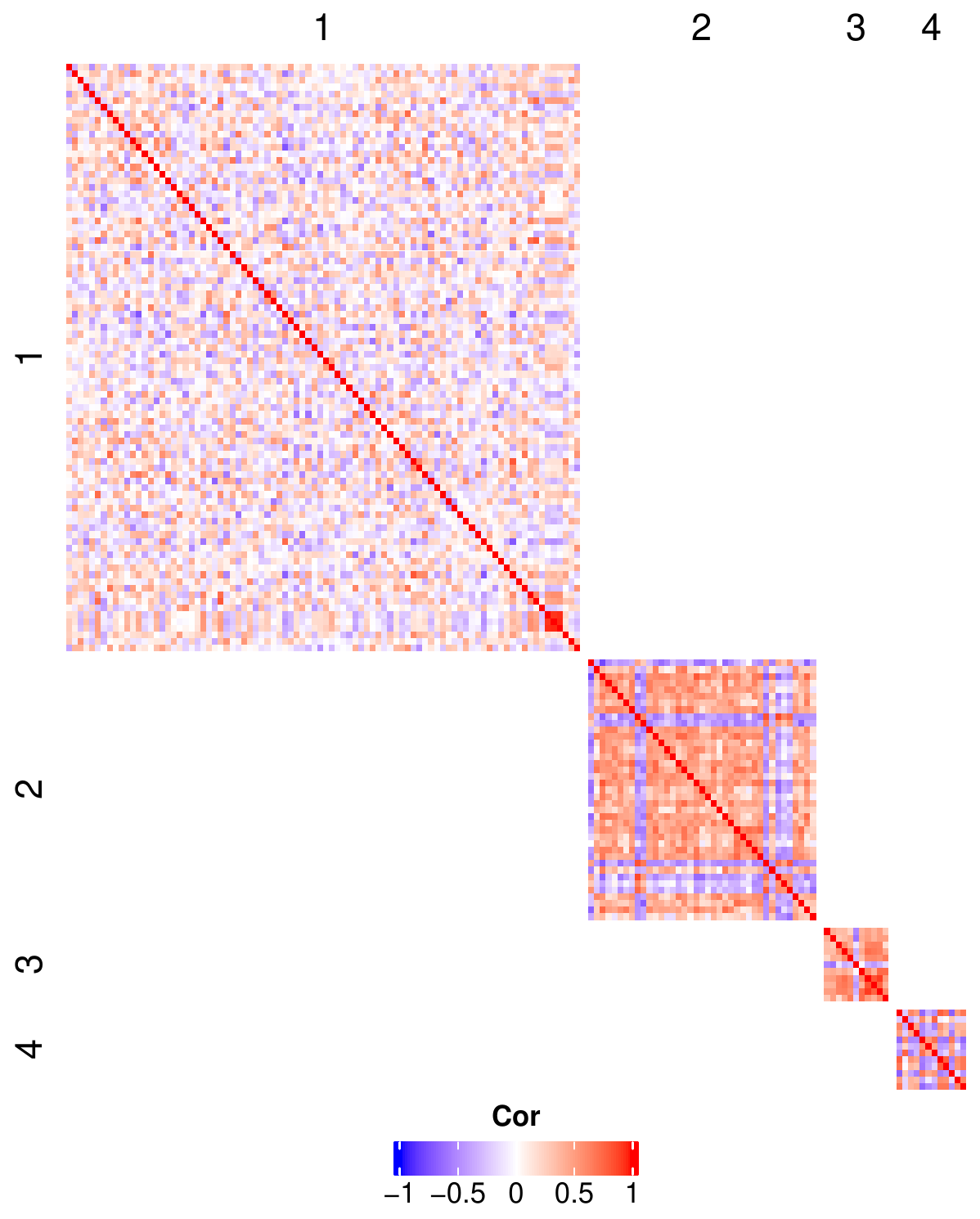}
\caption{Estimated correlation of predicted Group 2}
\end{subfigure}
\caption{True and estimated correlation matrices of the predicted groups from the \texttt{Golub}$_{150}$ dataset. Note that all 11 ALL samples were classified as Group 1 and 26 out of 27 AML samples were classified as Group 2.}\label{golub}
\end{figure}

In Figure \ref{golub}, we visualize the observed and estimated correlation structures in the two predicted clusters of the \texttt{Golub}$_{150}$ dataset. Column cluster 3 in Group 1 (i.e., the group comprising of all ALL samples) consists of 27 variables which have high correlations (either positive or negative) in the ALL samples. Out of the 27 variables, 24 of them were assigned to column cluster 1 in Group 2 (the group comprising of mostly AML samples). As evident from Figure \ref{golub}, the magnitude of the correlations among the variables in column cluster 1 of Group  2 is lower compared to that of column cluster 3 of Group 1. Interestingly, pathway analysis using Reactome \citep{reactome} identifies that 9 out of the above 24 genes are known to be involved in pathways related to platelet activation, signaling and aggregation. Platelets are specialized blood cells which are primarily responsible for preventing bleeding but are also recently found to promote tumor growth and metastasis \citep{yan2016role}. Furthermore, cancer cells have been shown to induce platelet activation and aggregation in several cancers, a phenomenon known as tumor cell–induced platelet aggregation \citep{jurasz2004}. Investigating the differences in correlation structures among these genes between AML and ALL samples could provide valuable biological insight.

\subsubsection{Bi-clustering in Topic Modelling}

Topic Modelling \cite{wallach2006topic} is an area of Natural Language Processing that tries to find abstract human-interpretable topics in a collection of an arbitrary number of documents. Essentially, the problem of grouping documents into unsupervised topics can be solved by leveraging clustering techniques. Human language is incredibly complex. There are multiple possible groupings one can find in a set of documents, some having a stronger signals-no-noise ratio than others. e.g., Trip advisor's hotel reviews can be grouped by review sentiment (recommend/do not recommend), hotel geographical location or more generalized hotel features (has access to the beach, superb breakfast, polite managers etc.). Each of the clustering can be desired depending on the application. However, classical clustering algorithms will require additional data post-processing to provide interpretations of the topics they produce \cite{sievert2014ldavis}.

%However, how do we tell the model which type of grouping we aim to obtain?
% One of the promising solutions to the above issues is bi-clustering. 

In this subsection, we apply the proposed bi-clustering algorithm to the topic modelling problem.
We test our algorithm on two real-world datasets of customer reviews available on Kaggle. The first is Trip Advisor hotel \cite{barkha_bansal_2018} data that consists of 20491 unique free-text hotel reviews accompanied by a numerical rating (from 1 to 5). The second one includes reviews and ratings of three Disneyland \citep{Chillar_2022} branches. We use only a subset of the data that reviews Disneyland Paris (13630 free-text responses). The language of reviews in both datasets is English. There are no reference labels available in this application. We aim to demonstrate that the proposed algorithm as a tool for summarizing customer reviews data.

For better explainability of the resulted topics and column groupings, we decided to apply the bi-clustering algorithm to the document-terms matrix instead of document-embeddings \cite{bertopic} data type. Testing the proposed algorithm with semantic text embeddings is left for future work. 

The entries of the document-terms matrix are term frequency-inverse document frequency scores (TF-IDF) \cite{aizawa2003information}. Let us define $\text{tf}(t, d)$ as a relative frequency of the term $t$ in some document $d$. Then, the inverse document frequency of the term $t$ given a collection $D$ of $N$ documents is computed as $\text{idf(t, D)} = \log \frac{N}{|{d \in D: t \in d}|}$.

To simplify, the term frequency measures how frequent a term is for a given document. The inverse document frequency measures how rare this term is in the overall collection of documents. We define a TF-IDF score of a term $t$ int he document $d$ as $\text{tf-idf}(t, d) =  \text{tf}(t, d) \text{idf(t, D)}$. The score is maximized when a specific term is frequent in a given document but uncommon for other documents. The TF-IDF score of the stop words is close to zero, as, by definition, stop words have a high probability of being present in every document in the collection.  

Large collections of documents may include over a thousand unique words. Therefore, before fitting the bi-clustering model, we select only $1000$ features with the highest average $TF$ scores. Additionally, we utilize \texttt{SelectKBest} from the sklearn \cite{scikit-learn} Python library to select $400$ terms that are the most predictive of the review rating. 

%\textcolor{blue}{**Should this be moved before for other datasets ** We used an elbow method and a silhouette score metric to select the optimal number of row clusters for every dataset. The proposed algorithm also inputs the number of column groupings. We discovered that small values $k$ for the number of groups result in most columns being grouped in one cluster, with other clusters consisting of $2-3$ columns. This is expected with hierarchical clustering. Therefore, we suggest selecting $k > 15$ for the purposes of topic modelling. Additionally, we utilized a family of Gaussian models with a shared covariance matrix. This simplifies the interpretation of feature groupings. }

\begin{figure}[h!]
  \centering
{\includegraphics[width=1\linewidth]{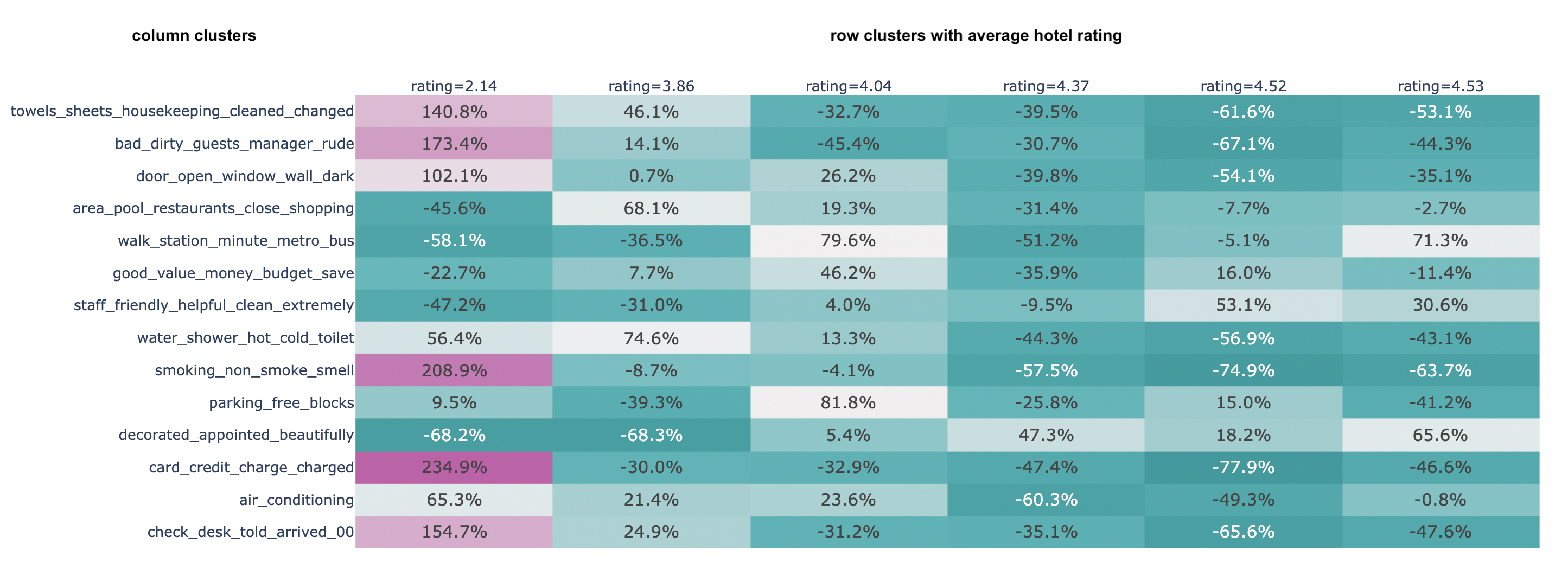}}\vspace{1mm}
  \caption{Resulting column and row clusters from the Trip Advisor hotel reviews. Entries of the heatmap represent how much ``more important'' a given column cluster is for a given row cluster than average. }
  \label{fig:topic_1}
\end{figure}

\begin{figure}[h!]
  \centering
{\includegraphics[width=1\linewidth]{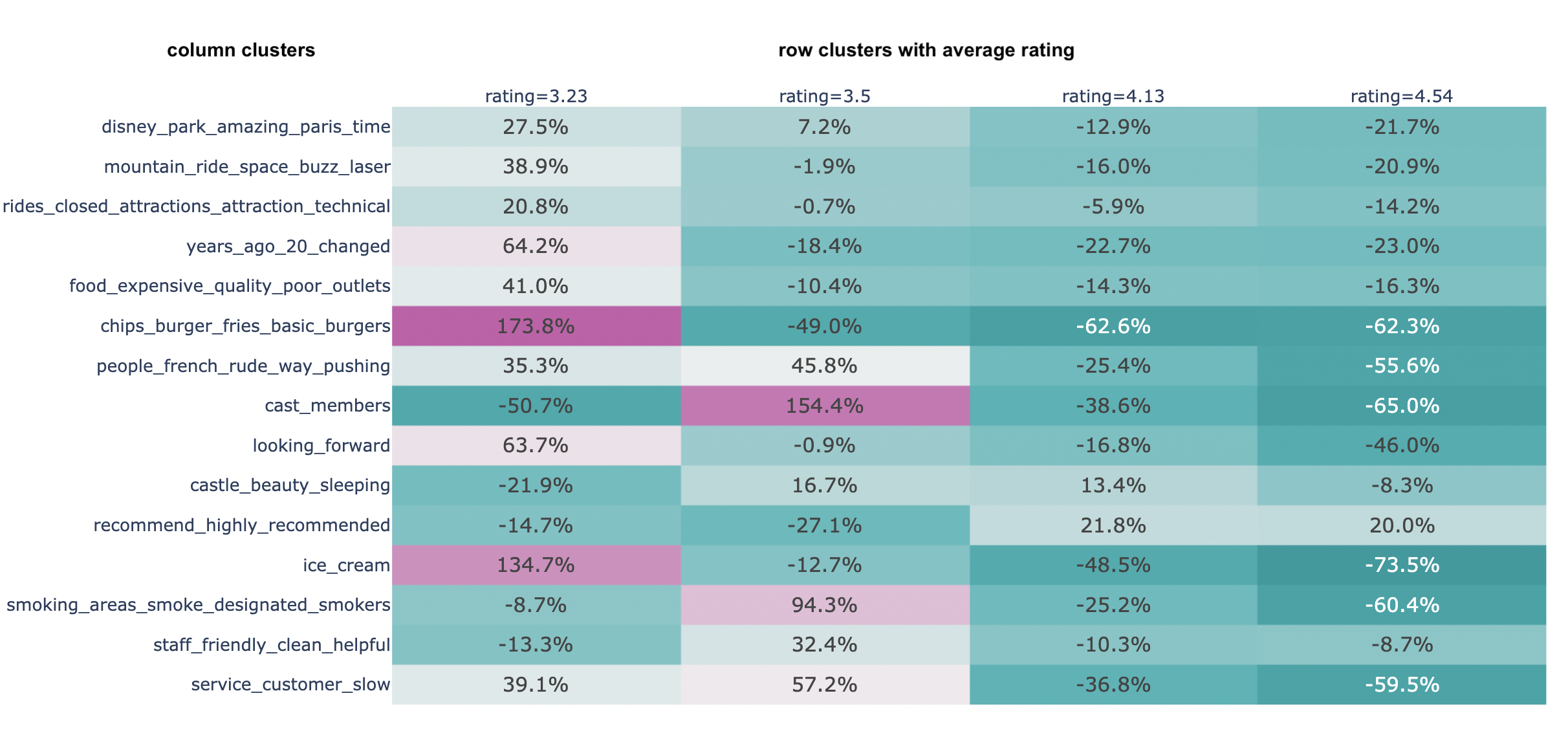}}\vspace{1mm}
  \caption{Resulting column and row clusters from the Disneyland Paris reviews. Entries of the heatmap represent how much ``more important'' a given column cluster is for a given row cluster than average. }
  \label{fig:topic_2}
\end{figure}

Figure \ref{fig:topic_1} and \ref{fig:topic_2} demonstrate the column and row groupings for the Trip Advisor and Disneyland Paris data. The heatmap entries are computed as the average TF-IDF score of terms inside the column grouping $A$ row cluster $B$ divided by average TF-IDF scores of $A$ among all row clusters and multiplied by $100$\%. This can be interpreted as how much "more important" a column cluster $A$ is for a row cluster $B$ than average. 

From Figure \ref{fig:topic_1}, we infer that lower hotel ratings may be attributed to the cigarette smell, rude managers, dirty conditions, problems with the check desk and unexpected credit card charges. The highest ratings are correlated with access to public transport or the hotel's location within walking distance of the city's landmarks, friendly hotel staff and beautiful decorations of the premises. Figure \ref{fig:topic_2} demonstrates that both feature clusters of food (ice cream and chips-burgers-fries) are highly associated with the lowest ratings of Disneyland Paris. It may signal that the food quality is lower than expected. At the same time, people are unhappy with closed attractions, rude staff, pushing and perhaps the fact that most people in Paris speak exclusively French.

\section{Conclusions}
%======================================================================
In this work, we have proposed a bi-clustering algorithm based on a finite mixture of multivariate Gaussian models with block diagonal covariance structure. We demonstrated that our approach has a comparable clustering performance to the state-of-the-art algorithms in the field while providing substantial improvement in the computational time. We also illustrate the applicability of the proposed approach on real world datasets from various fields.

%\bibliographystyle{agsm}
%\bibliography{biclustering}
\newpage

\appendix

\section{Standard Deviations for recovering the block-diagonal covariance matrix  }
\label{appendix-a}

\begin{figure}[h!]
  \centering
  \includegraphics[width=1\linewidth]{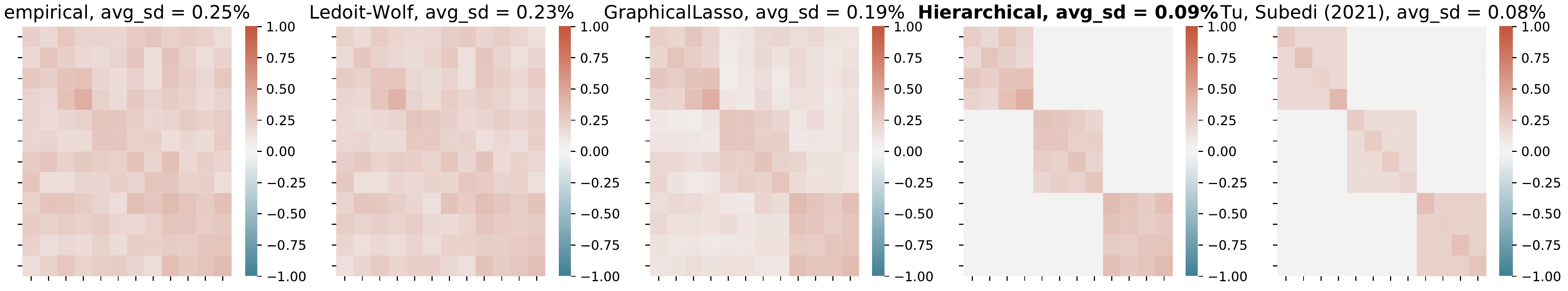} 
\includegraphics[width=1\linewidth]{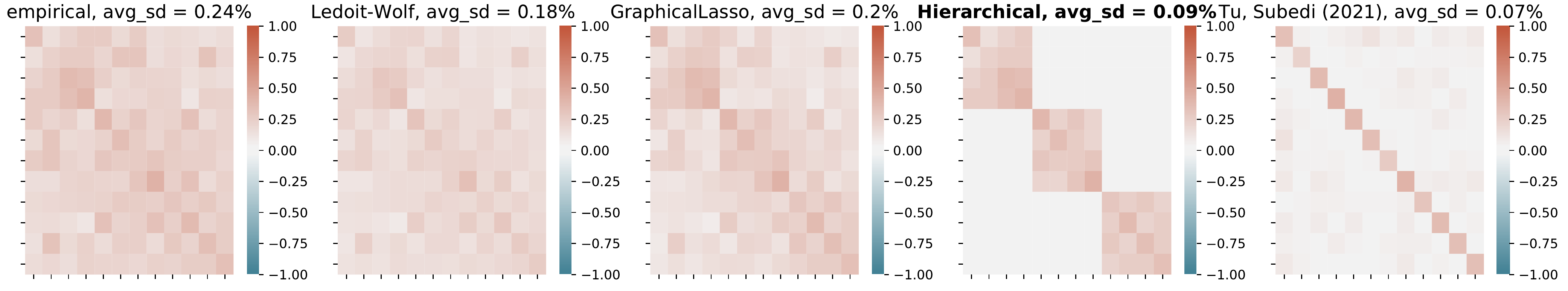}
  \caption{Standard deviations over ten runs of recovering block-diagonal covariance matrix with only positive (\textit{top}) / negative (\textit{bottom}) off-diagonal entries with five methods including MLE (empirical), Ledoit-Wolf, Graphical Lasso, Hierarchical (proposed) \& UCUU method from Tu, Subebi 2021 \cite{Tu2022AFO}}
  \label{fig:comparison-stds}
\end{figure}

\section{Accuracy for selecting the number of components}
\label{accuracy-plots}
 
\begin{figure}[h!]
\centering
\includegraphics[width=0.48\linewidth]{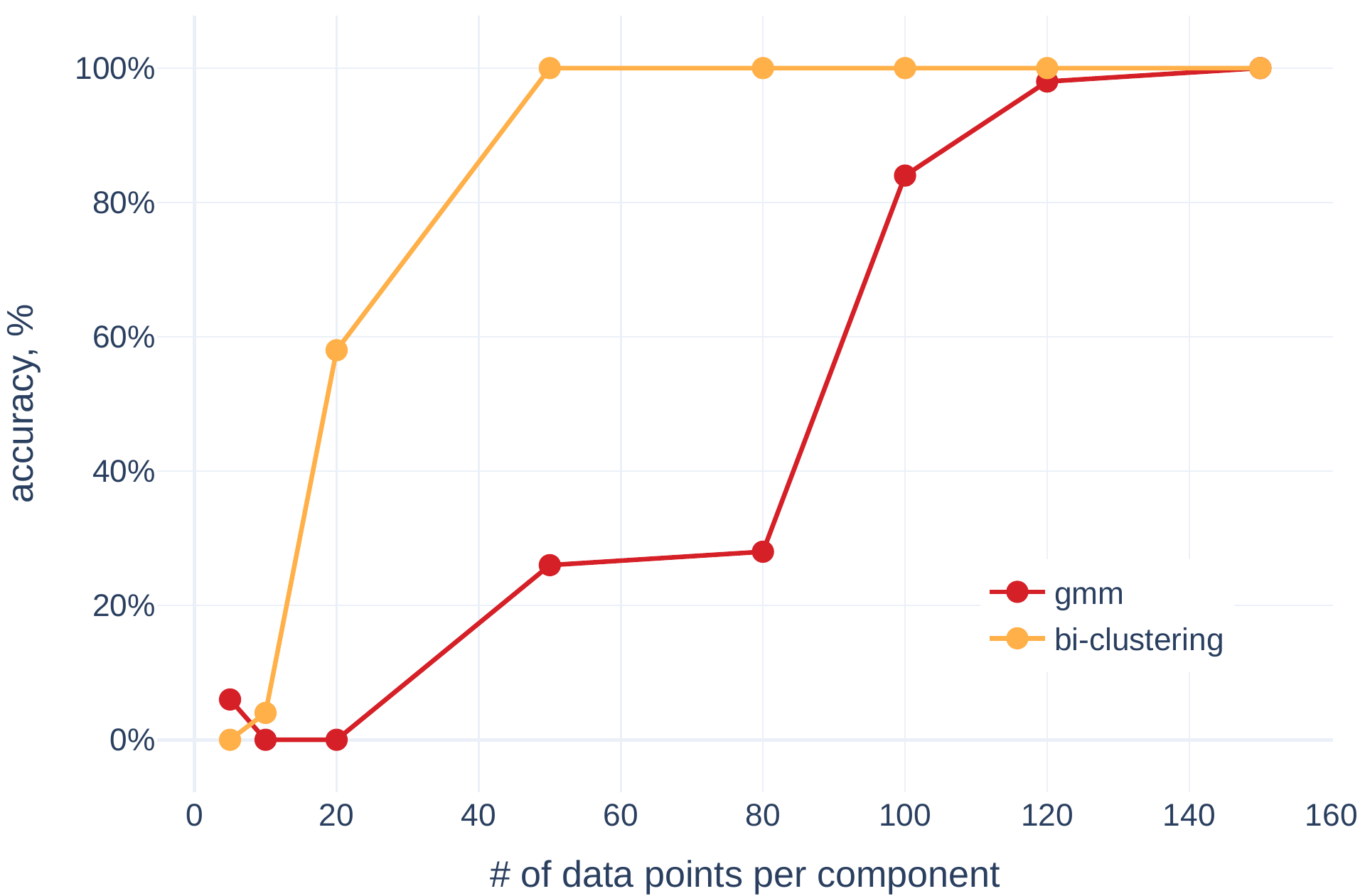}
\includegraphics[width=0.4\linewidth]{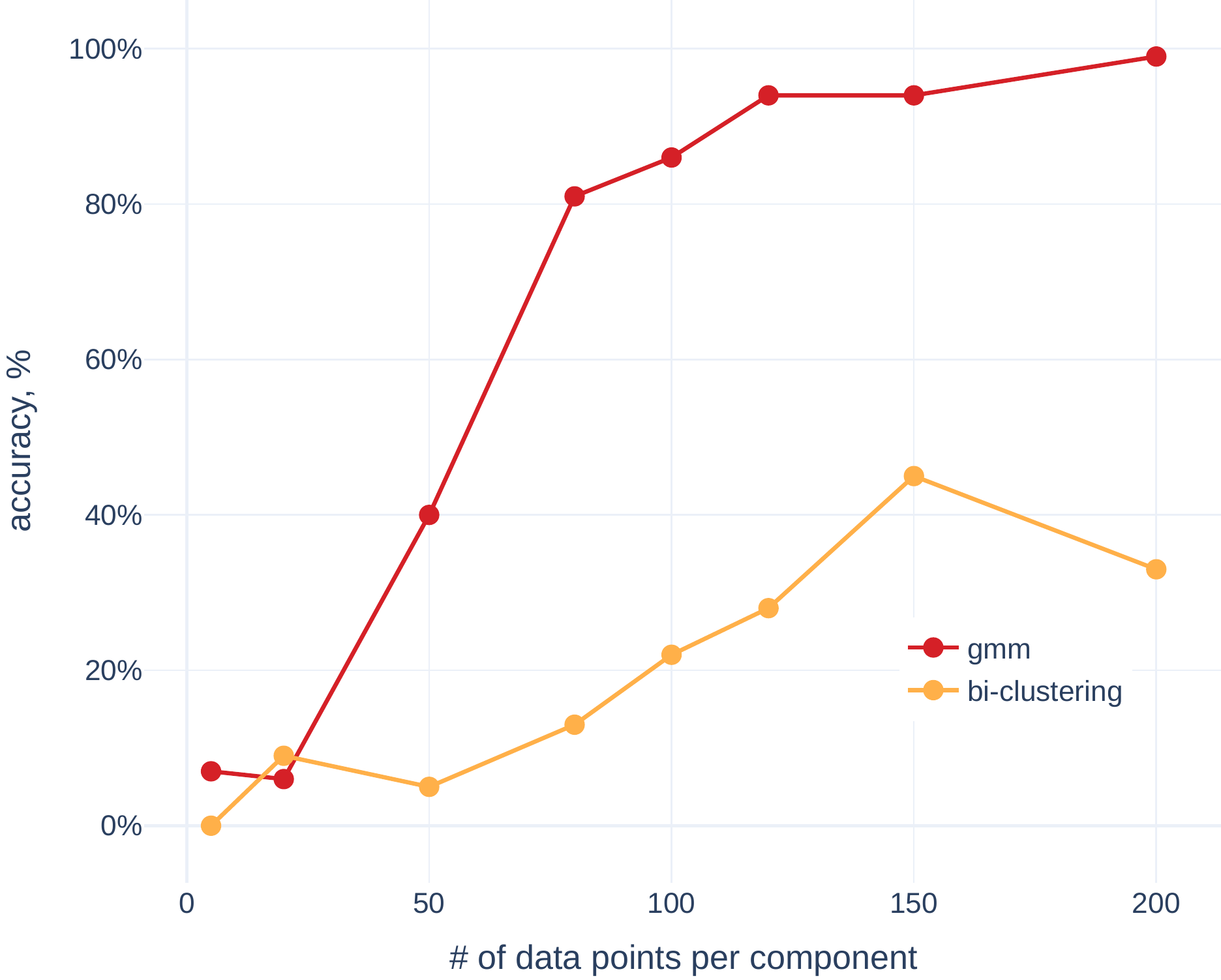}
% \captionsetup{justification=centering,margin=2cm}
\caption{Accuracy of the number of groups when the covariance structure is block diagonal (\textit{Left}) and a random positive definite matrix (\textit{Right}) for the GMM and the proposed bi-clustering model when $G=3$ while varying the number of observations within each component.
}
\label{fig:num-clusters-acc}
\end{figure}

\end{document}